**Contribution of chromosome-naïve transposon-seeded network of DNA templates to intracellular pools of protein dimers of liquid-liquid phase-separated condensate-forming proteins.**

Gennadi V. Glinsky[1]

[1] Institute of Engineering in Medicine
University of California, San Diego
9500 Gilman Dr. MC 0435
La Jolla, CA 92093-0435, USA
Correspondence:
genlighttech@gmail.com

**Abstract**

Molecular mechanisms governing protein-protein interactions (PPI) contributing to the biogenesis of membraneless condensates remain poorly understood. Observations reported in this contribution demonstrate that liquid-liquid phase-separated (LLPS) condensate-forming proteins and their variants harboring condensate-disrupting human pathogenic mutations are significantly enriched among PPI cascade constituents of the Hybrid Assembly Pathway (HAP) of endogenous multiprotein complexes (EMC). In contrast, no structured or disordered proteins defined as condensate-negative molecules were detected among PPI cascade constituents. Cumulative evidence derived from different lines of analytical inquiries support the concept of the significant contributions of PPI cascade constituents of HAP to the biogenesis of intracellular pools of cell type-specific protein heterodimers assembled with the engagement of at least one protein monomer selected from condensate-forming protein molecules. Analytical evidence documented in this study support the concept that genome-wide chromosome-naïve networks of transposon-seeded DNA-templates nucleate the initiation of PPI cascade chain reactions with the capacity to facilitate the concurrent iterative assembly in human cells of both the membraneless biomolecular condensates and the structured multiprotein complexes. AlphaFold 3D modelling of mutation patterns and analysis of partner-bound experimental structures identified variants located at a measured protein dimerization interface. This contribution nominates numerous protein heterodimers of potential relevance to human pathophysiology for follow-up mechanistic studies of potential fundamental and translational impacts.

## Introduction

DNA sequences derived from transposable elements (TEs) constitute approximately 50% of the human genome, providing major contributions to a multitude of structural features and regulatory functions at different levels of genomic and molecular organization of human cells (Lu et al., 2014; Elbarbary et al., 2016; Sundaram and Wysocka, 2020; Hoyt et al., 2022; Rosspopoff et al., 2025). Exceedingly large numbers and immense molecular diversity of TE-encoded DNA loci possess significant challenges for experimental explorations of their biological functions. Results of experimental studies focused on individual TE families facilitated the conceptual advances in understanding the regulatory and pathophysiological features that are common for DNA loci derived from different TE families (Glinsky, 2016-2026; and references therein). For example, based on their experimentally-documented regulatory functions during human preimplantation embryogenesis, DNA sequences of Long Terminal Repeats (LTRs) serving prior to the viral endogenization as evolutionary-ancient promoters of distinct families of Human Endogenous Retroviruses (HERVs), namely LTR7/HERVH; LTR5_Hs/HERVK; MLT2A1/HERVL; and MLT2A2/HERVL, were classified as human embryo regulatory LTRs (Glinsky, 2016-2026). Consistent with their regulatory functions in the early stages of human embryogenesis, TE families SVA, HERVK, and HERVH harbor markers of chromatin accessibility DNase I hypersensitive sites and are highly expressed in early embryos, but not in differentiated tissues (Gao et al., 2018; Grow et al., 2015; Guo et al., 2014). Similar to SVA, Alu transposons have open chromatin states in early human embryos and the number of expressed Alu appears higher if early embryos than in differentiated tissues (Gao et al., 2018). Genetic and epigenetic manipulations of human blastoids, a 3D model of the blastocyst, revealed a pervasive cis-regulatory effects of LTR5_Hs elements on the hominoid-specific diversification of the epiblast transcriptome in human blastoids (Fueyo et al., 2025). These human-specific regulatory mechanisms operating during pre-implantation embryogenesis appear linked to the many human-specific LTR5_Hs insertions which are unique to Homo sapiens (Fueyo et al., 2025). Reprogramming of differentiated somatic cells into induced pluripotent stem cells activates mobility of LINE-1 family of TEs (Wissing et al., 2012) and triggers retrotransposition of endogenous LINE-1 and Alu TE families in human induced pluripotent stem cells (Klawitter et al., 2016). Hundreds of thousands of LINE-1 family of TEs are primate specific, and thousands are human specific (Konkel et al., 2020). Thousands of hominoid-specific LINE-1 integrants are expressed in human induced pluripotent stem cells and cerebral organoids (Adami et al., 2025). Targeted CRISPR interference (CRISPRi) –mediated silencing of LINE1 retrotransposons demonstrated that the LINE-1 loci exert cis-acting transcriptional control of primate- and human-specific transcripts in human induced pluripotent stem cells and contribute to regulation of the early stages brain development (Adami et al., 2025). Consistently, it has been reported that: i) LINE-1 loci are dynamically expressed during neural differentiation; ii) LINE-1 activity is required for maintenance of neural progenitor pools in mice and humans; iii) LINE-1 loci play a critical role in both human and mouse brain development by regulating the rate of neural differentiation in a retrotransposition-independent manner (Toda et al., 2024).

Overall, the experimental observations reported by multiple independent groups of investigators demonstrated the crucial regulatory roles of the transcriptional activation of different families of TEs in human embryogenesis and the early stages of human brain development. These essential for human embryonic development TE-encoded regulatory loci are located in the accessible chromatin regions that became silenced in differentiated cells. However, recently presented findings support the concept postulating important regulatory roles of TE-derived DNA sequences residing within transcriptionally silenced heterochromatin regions in the broad range of physiological functions required for homeostasis maintenance of a cell.

Several reports documented multiple lines of observations revealing the initiation mechanism of protein-protein interactions (PPI) of hybrid assembly pathways (HAP) of structured endogenous multiprotein complexes, EMC (Glinsky, 2026a; b; c). According to HAP concept, chromosome-naïve networks of DNA sequences derived from different TE families function as initiation templates of PPI chain reactions which continually creates cell type-specific pools of hetero- and homodimers. Upon release from DNA templates, protein dimerization products serve as on-demand attractors of specific sets of monomers for the EMC assembly completion. Identification and initial characterization of key constituents of HAP of EMC (Glinsky, 2026 a; b; c) revealed a set of ~700 proteins designated initiator transcription factors (TF). Initiator TFs bind to double-stranded DNA templates in a sequence-specific manner. Upon binding to TE-seeded DNA templates, initiator TFs are poised to initiate PPI cascade chain reactions creating cell type-specific pools of protein homo- and heterodimers (Glinsky, 2026 a; b; c). Multiple lines of analytical evidence implicate approximately 4.5 million TE-encoded DNA loci as a massively parallel, protein-limited initiation layer for hybrid assembly pathways that continually nucleate PPI cascades in real time (Glinsky, 2026 a; b; c). These cascades congruently bifurcate into TF-TF PPI network branch, enriched in transcriptional programs defining neuroanatomical brain regions, and TF-HUB protein PPI network branch, enriched in constituents of synaptogenesis and synaptic architecture. This provides a conceptually unified genomic framework for the coordinated development and maintenance of brain regions and their connectivity (Glinsky, 2026 a; b; c).

Thus, according to the HAP concept, upon binding to TE-seeded DNA, initiator TFs activate the assembly of EMCs by iteratively engaging PPI cascade branches and creating cell type-specific pools of protein dimers. Most recently, molecular features and inferred biological functions of PPI cascade constituents were elucidated within the intracellular contexts of cell type-specific protein monomer pools documented in the Human Protein Atlas (HPA) of 154 human cell lines (Glinsky, 2016c). Within the contexts of HPA-defined intracellular proteomes, the four canonical families of PPI cascade constituents manifest the following key features of the molecular architecture and functional performance: a) the ubiquitous expression; b) balanced representation; c) the universal availability and engagements; d) constitutive {invariant) and cell type-specific assembly profiles of structured EMCs. The biological relevance of the PPI cascades has been demonstrated by cell-type-resolved multiprotein complex-engagement analysis across the HPA of human body. This analysis faithfully recovers without supervision the intrinsic lineage structures of EMCs and lineage-specific master-regulator multiprotein assemblies (Glinsky, 2026c). However, the potential relevance, if any, the HAP of EMC and specifically the PPI cascade constituents to the structural-functional characteristics, biological functions, and pathophysiological impacts of membraneless biomolecular condensates remains to be elucidated.

Membraneless biomolecular condensates bring proteins and nucleic acids into spatially restricted assemblies without a bounding membrane. Their components do not all play the same part: some proteins can drive liquid-liquid phase separation (LLPS) under specified conditions, others participate as clients, and other components defined as regulators may alter assembly without becoming condensate constituents (Banani et al., 2017; Pintado-Grima et al., 2025). These assignments are context-dependent. A protein observed in a condensate is not necessarily its autonomous driver, and an interaction between two annotated proteins does not establish that they bind within that condensate. These distinctions matter when an interaction network is used to infer what a cell can assemble from a set of transcription factors.

Available annotations of LLPS condensates answer different questions about this process. Banani et al. (2022) assembled 3,941 putative condensate-forming proteins by combining immunofluorescence-derived and sequence-based predictions with curated phase-separation records; a protein could enter from any one of those sources. The same study catalogued pathogenic variants and mapped a subset onto candidate condensate-promoting sequence features. Neither a predicted propensity nor a feature-overlapping variant demonstrates an altered condensate in a cell. Databases that annotate scaffolds, clients and condensate membership add more specific evidence, but their roles and inclusion rules differ (Mészáros et al., 2020; Ning et al., 2020; Rostam et al., 2023). Pintado-Grima et al. (2025) reconciled several such sources into positive sets with driver and client roles and a separate panel of condensate-negative proteins comprising condensate-negative sets of structured proteins and disordered proteins. Their autonomous-driver curation excludes known protein- or nucleic-acid-partner-dependent cases. It therefore cannot, by itself, settle whether a pathway that depends on DNA templates and interacting proteins initiates a condensate. Nor is absence from a positive catalogue evidence that a protein cannot participate.

A setting in which to examine these distinctions is the interaction cascade initiated by transcription factors associated with TE-derived binding sites. The previously described panel comprises 716 initiator transcription factors (TFs) with binding sites at 600 LTR5_Hs-encoded loci, supported by analyses of other TE families (Glinsky, 2026a; b; c). Gene-set enrichment distinguished 545 principal initiator TFs from 171 complementary initiators. Further enrichment against transcription-factor–transcription-factor and PPI HUB-protein libraries identified a 245-factor TF–TF branch and a 350-protein HUB branch; these are partially overlapping sets within a unified 1,122-symbol cascade panel. Repeated enrichment links the initiators to proteins in the interaction libraries, but the resulting network is not a physical particle. The PPI cascade postulated capacity to support condensate initiation and multiprotein-complex assembly remains an inference about possible routes, not a direct experimental demonstration of either assembly in a living cell.

Three gaps separate that inference from a testable account of condensate engagement. First, the two interaction branches need not have the same relation to curated LLPS drivers, clients or putative negatives, and source labels can overlap with the evidence used to construct a network. Second, a protein–protein interaction catalogue records possible partners without specifying which are expressed together in a particular cell; protein availability and graph degree can create apparent enrichment in recurrent interaction pairs. Third, a compartment annotation, including one based on a related functional Gene Ontology term, does not locate an interaction in a phase-separated body. These issues require comparisons on explicitly mapped human proteins, cell-type-conditioned pair sets and null models that preserve relevant properties of the interaction graph. They also require keeping source-defined negatives distinct from proteins that are merely unannotated: the negative sets of Pintado-Grima et al. (2025) were screened against known LLPS entries and their first-degree BioGRID interactors, a selection that may affect their representation in a PPI-derived panel without making overlap impossible.

Investigations of pathogenic-variant labels introduce a related but separate question. A long or heavily studied protein can carry many catalogued labels, and several substitutions at one residue can make a structural display look more densely marked than its distinct affected-site count warrants. Local concentrations of sites along the sequence need not imply additional proximity in a predicted three-dimensional fold. Conversely, a wild-type AlphaFold model can locate a reference residue without predicting the conformational or assembly effect of its mutation. Quantitative comparison therefore needs to separate length-adjusted catalogue burden, distinct-site concentration along the sequence and proximity in the reference structure. Each has a different null and a different biological limit.

The simplest definition of biomolecular condensates could be articulated as intracellular membraneless compartments that segregate macromolecular complexes contributing to various biochemical reactions and biological processes. The typical DNA molecules that store and relay genetic information exist as chromatin fibers which are defined at the most fundamental, ground-state level as the complexes of DNA and histone proteins. Chromatin fibers undergo phase separation into biomolecular condensates in vitro and in vivo and DNA linker lengths influence the chromatin condensate properties. The material properties and stability of chromatin condensates exemplified by distinct viscous and elastic behaviors versus purely viscous behaviors varied between strongly or weekly connected networks of chromatin fibers with different lengths of DNA linkers (Zhou et al., 2025). Notably, the condensates of weekly connected chromatin fiber networks of relatively extended lengths of DNA linkers manifest purely viscous behaviors with more rapid molecular diffusion and ready dissolution by acetylation (Zhou et al., 2025). However, the potential molecular cross-talks between TE-encoded DNA template-nucleated PPI chain reactions and the PPI cascade constituents vis-a-vis the LLPS reactions and the assembly of biomolecular condensates have not been investigated.

In this study, the 1,122 initiator-associated PPI cascade constituents of TE-endowed chromosome-naïve HAP of EMCs (Glinsky, 2026a; b; c) were examined against source-defined protein constituents of condensate compartments, proteins with curated LLPS roles and LLPS condensate-negative protein sets, and pathogenic-variant annotations. The available identifiers were mapped to analyzed human genes and proteins, related PPI cascade branch-reached proteins and reconstructed intracellular pools of heterodimers to expression-defined availability across 154 human cell types, and compared annotation pairing with degree- and detection-breadth-matched nulls. Human pathogenic mutation-label burden were distinguished from sequence hotspots and determined whether those hotspots show additional proximity in AlphaFold wild-type models. These analyses test the annotation and structural predictions of the PPI cascade while retaining the distinction between inferred interaction capacity and observed condensate formation.

## Results

### Enrichment of putative phase-separated condensate-forming proteins among the PPI cascade constituents.

Of the 1,122 PPI cascade constituents (Glinsky, 2026a; b; c), 1,011 proteins are expressed in the Human Protein Atlas (HPA) of 154 human cell types and 409 proteins are on the list of the putative phase-separated condensate-forming proteins (Table 1). Annotation analysis of putative phase-separated condensate-forming proteins (Banani et al., 2022) tracks detection and study depth: 27.1% of genes measured in the HPA single-cell records of 154 human cell types are on the list against 19.1% of the human proteome at large (Table 1), while the PPI cascade network rate of putative condensate-forming proteins is 40.5%, which corresponds to the relative enrichment of 2.12 ($p = 1.4e-58$). Against the population the PPI cascade constituents are actually drawn from, the enrichment value of 1.49 is significantly distant from chance (Table 1). Thus, the number of putative condensate-forming proteins is significantly enriched among the PPI cascade network constituents (Figure 1).

**Table 1.** Enrichment analysis of the PPI cascade constituents* among the 3,862 proteins defined as the candidate protein components of phase-separated condensates.

| Background | Background rate | Network rate** | Ratio, O/E | Number, expected | P value |
|---|---|---|---|---|---|
| **Human proteome (20,190)** | 19.1% (3862/20190) | 40.5% (409/1,011) | 2.12 | 193.4 | 1.4e-58 |
| **HPA***-measured genes (9,701)** | 27.1% | 40.5% | 1.49 | 273.6 | 1.0e-22 |
| **Proteome ∩ HPA-measured (9,635)** | 27.2% | 40.5% | 1.49 | 275.4 | 4.1e-22 |

Legend: *, 1,011 of the 1,122 HAP constituents map into the reviewed proteome; **, 409 of 1,011 HAP constituents overlap 3,862 condensate proteins; O/E, observed/expected ratio; HPA***, Human Protein Atlas; ∩ = intersection. 3,868 of the 3,941 entry names (Banani et al., 2022) carry a primary gene symbol, giving 3,862 distinct symbols of proteins defined as the candidate protein components of phase-separated condensates.

It was of interest to carry out the comparative analysis of the different families of PPI cascade constituents (Table 2; Figure 1). The PPI cascade families T245 (TF-TF PPI cascade branch) and H350 (TF-HUB protein PPI cascade branch) sit at median degree 488 and 506, reached in 66.5% and 62.0% of cell types (matched on both).

**Table 2.** Comparisons of analytical metrics of the PPI cascade constituents.

| PPI cascade subset | n | condensate % | z | median degree | % library terms | % reached | tftf / hub reach | % available |
|---|---|---|---|---|---|---|---|---|
| **716 initiators** | 716 | 39.8 | +3.94 | 178 | 25.4 | 15.1 | 14.7 / 2.8 | 56.1 |
| **T245 TF–TF family** | 245 | 47.8 | +3.22 | 488 | 100 | 66.5 | 66.1 / 11.8 | 80.0 |
| **P545 principal initiators** | 545 | 40.7 | +3.01 | 249 | 27.9 | 16.5 | 16.1 / 2.9 | 55.8 |
| **C171 initiators complementary** | 171 | 36.8 | +0.53 | 30 | 17.5 | 11.1 | 9.9 / 2.3 | 56.6 |
| **128 proteinopathy-bound targets** | 128 | 36.7 | −1.36 | 586 | 100 | 85.9 | 25.8 / 68.8 | 92.1 |
| **81 SynGO hub proteins** | 81 | 29.6 | −2.53 | 580 | 100 | 93.8 | 3.7 / 93.8 | 95.1 |
| **H350 hub family** | 350 | 26.9 | −5.77 | 506 | 100 | 62.0 | 8.0 / 62.0 | 92.7 |

However, they reached through opposite libraries: T245 PPI cascade branch reached 66.1% TF-TF library and 11.8% TF-HUB library; while H350 PPI cascade branch reached 8.0% TF-TF library and 62.0% TF-HUB library. Importantly, they are carrying significantly different phase-separated condensate annotation of 47.8% (T245 branch) against 26.9% (H350 branch). Thus, T245 family of the PPI cascade constituents is fully a library term, reached in two thirds of cell types, and condensate-enriched at z +3.22 (Table 2). The PPI cascade plateau does not dilute the condensate content — seed layer 39.8%, network 36.5%, TF-TF branch at plateau 52.1%, TF-HUB branch 38.8%, all reached terms 42.9%. Thus, the TF-TF PPI cascade branch (T245) walk into the condensate-rich engagement (Figure 1). Results of the above analyses appear to license to call the TF-TF PPI cascade branch a phase-separated condensate arm of the PPI cascade network.

During the analyses of the 51,439 retained, cell-reconstructed PPI cascade heterodimers identified in 154 human cell types, it has been observed that 2,440 of 3,862 putative condensate-forming proteins (63.2%) participate as monomers in formation of at least one of heterodimers, while 1,422 candidate cascade-forming proteins (36.8%) did not participate as monomer subunits in PPI cascade-associated intracellular heterodimer pools. In human cells, the proportion of candidate condensate-forming proteins participating in PPI cascade-initiated heterodimerization appears significantly higher compared to non-candidate condensate-forming proteins. Within the 2,529 eligible candidate condensate-forming proteins (having at least one term–member protein-protein interaction in the original Enrichr TF–TF PPI or PPI HUB libraries), 2,440 (96.48%) participate in at least one cell type-reconstructed retained heterodimer. Of the 6,784 HPA-measured, raw-library-eligible non-candidate condensate-forming proteins, 6,433 (94.83%) participate in at least one cell type-reconstructed retained heterodimer. A direct one-sided Fisher's exact test of those 2 eligibility groups gives odds ratio (OR) 1.50, $p = 0.000363$. However, this difference between 2 eligibility groups may be at least in part due to the degree-, expression- and cascade-anchor-matched expectation distinctions (data not shown).

Unsupervised hierarchical clustering analysis revealed biologically-meaningful similarity association matrices of human cells based on common intracellular condensate-forming protein heterodimer pools, assembly of which was inferred from the PPI cascade chain reactions nucleated at TE-encoded DNA templates (Figure 2). Informative examples of unsupervised hierarchical clustering of human cells reflecting their common developmental, anatomical, and biological features include a human spermatogenesis cluster (6 cell types); a cluster of cell types residing in the human brain (10 cell types); a cluster of 8 cell types including pituitary gland cells (5 cell types) and retina cells (3 cell types); a cluster of 14 cell types segregating immune, lymphoid, and vasculature cells; a hematopoiesis cluster of 7 cell types, including hematopoietic stem cells, 5 types of progenitor cells of white and red blood cells, and thymocytes (Figure 2). In the Table 3 top-scoring examples are reported of diverse representations of the PPI cascade constituents-associated protein heterodimers with at least one monomer protein selected from candidate condensate-forming proteins of five groups of human cells defined by the unsupervised hierarchical clustering of 154 human cell types. Surveying the clustering patterns along the dendrogram tree identify multiple additional examples of the different size groupings of cell types reflecting the commonality of developmental, anatomical, and biological features of segregated cell types.

Based on reported herein observations, it has been concluded that cumulative evidence derived from different lines of analytical inquiries support the concept of the significant contributions of TE-endowed chromosome-naïve networks of DNA templates to the biogenesis of intracellular pools of protein heterodimers assembled with the engagement of at least one protein monomer selected from candidate condensate-forming protein molecules.

**The PPI cascade network has the capacity to the concurrent initiation in human cells of multiprotein assembly lines of the phase-separated condensates and the CORUM multiprotein complexes.**

Given the essentially unlimited DNA templates, it could be predicted that the PPI cascade network has the capacity to run the condensate initiation arm (via TF-TF branch; T245) and the initiation of fully structured multiprotein complexes assembly (via TF-HUB protein branch; H350). The expectation holds in both directions of condensate-depleted and complex-enriched claims. A test of CORUM membership was carried out: same degree-matched null, 20,000 draws demonstrated the enrichment of multiprotein complexes claim (Table 4). The PPI cascade hub branch H350 is CORUM-richer than the TT-TF PPI branch T245 in 154 of 154 cell types (median +8.0 points), and the TF-TF PPI cascade branch is condensate-richer in 154 of 154 cell types (median +12.9).

Across the panel of 154 human cell types, there were no exceptions either way, so the two-arm direction is not a summary artefact. But the families' members themselves do not carry it. On CORUM membership of multiprotein complexes, T245 (46.1%) and H350 (44.6%) are indistinguishable, and both only modestly above a degree-matched draw (z +1.86, p 0.080; z +2.07, p 0.049). The 716 initiators are CORUM membership-depleted, 14.5% against 43.8% in the rest of the network, z −4.20. So the apparent CORUM complexes/condensate members split is a property of what each branch reaches, not of the seed or partner family property as a set.

**Table 4.** At the plateau the branches separate in both labels, and in the same direction in every cell type (medians over 154 cell types).

| PPI cascade stage | CORUM % | condensate % | complexes per member |
|---|---|---|---|
| **seed layer, 716 initiators** | 14.5 | 39.8 | 0 |
| **network 1122** | 25.1 | 36.5 | — |
| **tftf branch at plateau** | 57.7 | **52.1** | 1 |
| **hub branch at plateau** | **65.8** | 38.8 | 2 |

However, the constituents of the PPI cascade network could not be viewed as separate populations both in structural and functional terms. Over the 1,122-protein PPI cascade network: 122 proteins carry designations of both CORUM complexes and putative condensate-forming proteins, 160 proteins are CORUM only, 287 proteins are condensate-forming members only, while 553 PPI cascade constituents neither. Thus, a total of 43.3% of the PPI cascade network's CORUM members are also condensate-forming candidates, and the two associate positively rather than exclusively (OR 1.47, Fisher p 0.0066). Any two-arm dichotomy description is about the relative enrichment, not about disjoint populations, because 38.8% of the PPI cascade hub branch plateau is condensate-annotated and 57.7% of the TF-TF branch plateau is in CORUM multiprotein complexes (Table 4). These observations reflect the PPI cascade constituents' enrichment, not partition.

In summary, considering the essentially unlimited availability of DNA templates, the PPI cascade network has the capacity to run both the condensate arm and the CORUM arm with no anticipated competition: both reach a plateau in every cell type, from the same 716 seeds of initiator TFs. The PPI cascade TF-TF branch's plateau is richer in the condensate-forming proteins and the PPI cascade TF-HUB branch's plateau is richer in the CORUM multiprotein complexes, consistently, in 154 of 154 human cell types. However, the T245 and H350 families are not the two functionally-segregated PPI cascade arms, or the separate populations because both of the families manifest cross-annotation profiles: 38.8% of the PPI cascade H350 plateau is condensate-annotated and 57.7% of the PPI cascade T245 plateau is in CORUM multiprotein complexes.

**Assignment of the PPI cascade constituents to the 35 specifically defined phase-separated condensate compartments.**

It was of interest to extend the analytical experiments to resolve the 35 specifically identified phase-separated condensate compartments (Banani et al., 2022) by assignments to the 4 families of the PPI cascade constituents within the proteome contexts of the 154 human cell types. Reconstruction of 154 cell-resolved dimer repertoires from public HPA data demonstrate that among annotations with ≥10 both-annotated pairs and ≥5 expected pairs in all 154 cell types, the largest median observed/expected ratios are: PML body 4.67, DNA repair focus proxy 3.76, chromatin 3.01, transcriptional condensate proxy 1.92, focal adhesion 1.79, centrosome 1.70, nucleolus 1.57, nuclear body 1.49. These are tendencies for annotated proteins to consistently pair within an intracellular candidate dimer repertoire, which is consistent with (but not an experimental proof of) that the dimers occupy and may form those compartments.

The formal aggregate test across 154 cells finds 10 of 35 compartment annotations overrepresented for pairs with both endpoints annotated after BH correction (Figure 3; Table 5). These tests are asking the specific question: do proteins assigned to the same 35 condensate compartments co-occur as interacting pairs more often than expected from their frequency among proteins with similar graph degree and across-cell expression breadth? The null uses 100,000 joint annotation permutations matched on dimer-graph degree and HPA detection breadth; it keeps the interaction graph and each pair's cell-occurrence count fixed, so cells are not treated as independent replicates. PML body, chromatin, and the source's DNA-repair-focus and transcriptional-condensate GO proxies each reached the permutation resolution floor ($p = 1/100{,}001$; $q = 8.75 \times 10^{-5}$). Cajal body, centrosome, nuclear body, DNA-replication-condensate proxy, nucleolus, and focal adhesion also passed $q < 0.05$. The same ten pass when each distinct pair is counted once rather than weighted by its cell frequency. However, an experimentally supported PPI between two GO-annotated proteins should not be interpreted as the evidence that this specific dimer forms in the named compartment or that universal presence of a specific phase-separated structure across the panel of 154 human cell types is the evidence that the specific condensate compartment is necessary or required for cell survival.

**Table 5.** Statistical metrics of 10 overrepresented phase-separated condensate compartment annotations.

| Phase-separation Compartment | Annotated endpoints | Observed cell–pair occurrences | Null mean | Observed/null | Permutation p | BH q (35) |
|---|---|---|---|---|---|---|
| **DNA repair focus†** | 152 | 28,018 | 6,951 | 4.03 | $1.0\times10^{-5}$ | $8.75\times10^{-5}$ |
| **PML body** | 56 | 14,243 | 2,649 | 5.38 | $1.0\times10^{-5}$ | $8.75\times10^{-5}$ |
| **chromatin** | 264 | 170,020 | 53,315 | 3.19 | $1.0\times10^{-5}$ | $8.75\times10^{-5}$ |
| **transcriptional condensate†** | 807 | 562,067 | 270,864 | 2.08 | $1.0\times10^{-5}$ | $8.75\times10^{-5}$ |
| **Cajal body** | 24 | 1,254 | 77 | 16.27 | $4.00\times10^{-5}$ | $2.80\times10^{-4}$ |
| **centrosome** | 344 | 46,878 | 25,438 | 1.84 | 0.00127 | 0.00741 |
| **nuclear body** | 504 | 110,156 | 68,819 | 1.60 | 0.00166 | 0.00830 |
| **DNA replication condensate†** | 47 | 1,643 | 294 | 5.59 | 0.00217 | 0.00914 |
| **nucleolus** | 557 | 103,235 | 65,101 | 1.59 | 0.00245 | 0.00914 |
| **focal adhesion** | 316 | 71,423 | 43,509 | 1.64 | 0.00261 | 0.00914 |

Legend: table entries report statistical metrics associated with 10 overrepresented LLPS compartments selected from a total of 35 LLPS compartment annotations described in Banani et al. (2022).

The reconstructed cell-specific dimer repertoires contain pairs whose two proteins share condensate-associated GO annotations. Their observed pairing exceeds a degree-bin-shuffled expectation for several annotations, consistent with the PPI cascade cascade supplying candidate interaction partners to condensate-associated protein pools. Furthermore, the PPI cascade selects cell-conditioned repertoires of experimentally supported interacting protein pairs whose partners are annotated to the same condensate-associated compartment. This is consistent with the PPI cascade potentially supplying interaction-capable protein dimers to those phase-separated condensate compartments.

In the reconstructed dimer repertoires, all 154 human cell types have observed/degree-bin-expected pairing ratios above 1 for PML body, chromatin, focal adhesion, centrosome, nucleolus, nuclear body, and the source's DNA-repair and transcriptional-condensate proxies (Figure 3). In contrast, the ribonucleoprotein granule exceeds 1 in only 21/154 cells. The very nature of the highlighted condensates and the cellular functions associated with the highlighted annotations make a shared, biologically important condensate-related repertoire a plausible interpretation of their universal representations across 154 human cell types. Transcription, chromatin organization, DNA repair, and ribosome biogenesis are not specific niche cell-type programs.

At nTPM ≥ 1, the mean per-cell fraction of the reached TF–TF PPI cascade branch plateau assigned to the source's transcriptional condensate functional GO proxy is 68.5%, versus 27.1% for the PPI cascade TF-HUB branch plateau (Figure 3). The TF–TF fraction is higher in 154/154 cells. Chromatin is 28.0% versus 16.8% (TF–TF higher in 154/154); focal adhesion is 1.26% versus 13.8%, and centrosome 3.80% versus 10.3% (PPI cascade hub branch higher in 154/154). Under the top-4,000 rule the corresponding means are 72.3% versus 28.6% (transcription proxy), 33.2% versus 17.8% (chromatin), 2.15% versus 16.5% (focal adhesion), and 3.89% versus 10.3% (centrosome).

**Identification of the candidate condensate-disrupting pathogenic mutations among the PPI cascade constituents**.

Previously reported analyses identified 2,367 human genes harboring ≥1 pathogenic mutations, 1,741 (74%) of which have ≥1 pathogenic mutations mapped to condensate-promoting features in genes encoding the candidate phase-separated condensate-forming proteins (Banani et al., 2022). The 1,011-protein PPI cascade network contains 409 unique mapped genes encoding putative condensate-forming proteins; 299 of those have a human pathogenic mutation and 269 (90%) have a condensate-disrupting pathogenic mutation in genes encoding condensate-forming proteins (Figure 4). To determine whether condensate-disrupting pathogenic mutation-harboring genes are enriched among distinct families of the PPI cascade constituents, the comparator is other candidate genes with ≥1 pathogenic mutations, not the entire proteome. Fisher tests compare values ascertained for each family with all other mutation-ascertained genes (Table 6).

Under the expression nTPM ≥ 1 rule, the median TF–TF PPI cascade plateau contains 65 condensate disruptive mutation-positive genes among 84 reached genes encoding candidate condensate-forming proteins (median per-cell fraction is 76.9%). The TF-HUB PPI cascade plateau medians are 54 among 84 (63.8%). Under the availability rule of top-4,000 expressed proteins, the corresponding medians are 42 among 53 (79.2%) and 41 among 62 (66.1%). In the 154 human cell types, these are annotations of wild-type genes with published pathogenic variants, not counts of variant-bearing proteins in those cells.

Notably, large fractions of the PPI cascade-supplied heterodimers have candidate condensate-disrupting mutations. Of 51,439 reconstructed heterodimer types, 22,546 (43.8%) have ≥1 candidate condensate-disrupting mutations and 3,325 (6.5%) have two. Across 5,919,637 cell–heterodimer occurrences, the corresponding incidences are 2,705,299 (45.7%) and 415,171 (7.01%). Among PML-body-annotated interacting pair types, 68/109 (62.4%) have ≥1 candidate condensate-disrupting mutations and 18/109 (16.5%) have two; the analogous counts are 978/1,269 (77.1%) and 334/1,269 (26.3%) for chromatin compartment; 109/196 (55.6%) and 21/196 (10.7%) for the DNA-repair-focus compartment GO proxy; and 3,096/4,827 (64.1%) and 800/4,827 (16.6%) for the transcriptional compartment GO proxy. Supplementary information (table banani_mutation_compartment_pair_overlay.csv) covers all 35 condensate compartment labels.

**Table 6.** Identification of the candidate condensate-disrupting pathogenic mutations among the PPI cascade constituents*.

| Set | Feature-overlap + / mutation-ascertained | Comparator + / mutation-ascertained | OR | BH q |
|---|---|---|---|---|
| **P545** | 154/168 | 1,587/2,199 | 4.24 | $4.45\times10^{-9}$ |
| **C171** | 37/40 | 1,704/2,327 | 4.51 | 0.00452 |
| **T245** | 90/96 | 1,651/2,271 | 5.63 | $6.40\times10^{-7}$ |
| **H350** | 57/70 | 1,684/2,297 | 1.60 | 0.168 |
| **All mapped network candidates** | 269/299 | 1,472/2,068 | 3.63 | $6.25\times10^{-13}$ |

Legend: *, Values are gene counts, not mutation-event counts. BH correction covers these four family tests plus the whole-network contrast (five tests).

These observations demonstrate that many genes encoding the PPI cascade constituents have pathogenic variants mapped to condensate protein-promoting features, and many experimentally supported PPI cascade candidates link such genes to phase-separated condensate compartment annotations. It is unknown whether those pathogenic variants affect the PPI cascade-supplied protein dimers or whether a named phase-separated structure fails as a consequence of those mutations.

It remains to be determined experimentally whether any of these pathogenic potentially condensate-disrupting mutations may affect the PPI cascade performance or change the stoichiometry of protein-protein interactions during the protein dimerization. However, there are at least two examples of the effects of putative condensate-disrupting mutations on self-association in at least two PPI cascade proteins (not established for their partner-specific cascade heterodimers). Alignments of the 12,237 condensate-disrupting variants in 269 mapped PPI cascade constituents-coding genes with the MID and low-complexity sequence (LCS) intervals identified 4,792 parseable single-residue missense changes: 3,251 lie in an annotated MID, 2,561 in an LCS and 1,020 in both. An MID can bind DNA, RNA, or a protein, so overlap with an MID is not a proof that any named PPI edge changes. The other 7,445 labels include indels, truncations, frameshifts and other mutation descriptions whose effects cannot be assigned by checking a single residue. Forty-four missense changes intersect an explicitly named oligomerization/tetramerisation domain, all in TP53 (Table 7).

The examples reported in the Table 7 show that a mutation of the PPI cascade protein can affect self-association resulting in either enhancement (STAT3 homodimers) or disruption/destabilization (p53 tetramers) of protein-protein interactions. These observations are of potential importance to human cells, because there are a total of 191 homodimers identified in 154 human cell types of the HPA panel and most of them appear relevant to membraneless biomolecular condensates, including all analyzed herein classifications of condensate-forming proteins. The 191 validated homodimer types yield 26,560 reconstructed cell–pair occurrences in 154 HPA cell types. Per-cell counts range 134–185 (median 174.0). The per-protein cell breadth is 2–154 (median 153.0); 76 homodimer-forming proteins occur in all 154 reconstructed intracellular protein dimer pools of human cells. All 191 homodimer-forming proteins are constituents of the HUB-proteins PPI cascade branch (H350 family). A total of 59 of homodimer-forming proteins harbor human pathogenic mutations. Among these, 20/45 sequence-testable genes have a detected cohort-adjusted missense-site hotspot. Examples with cohort-adjusted sequence hotspots include APP, HNRNPA1 and CREBBP genes. The analysis of wild-type AlphaFold models do not provide mutation-specific structures and cannot establish mutation-dependent changes of homodimer interfaces or effects on phase separation.

**Table 7.** Examples of the PPI cascade proteins TP53 and STAT3 condensate-disrupting mutations affecting self-association

| Source variant | Matched feature and tracked self-pair | Independent evidence | What it does not demonstrate |
|---|---|---|---|
| **TP53 Arg337His** | p53 tetramerisation domain (319–357); TP53–TP53 is in the union in 152/154 cell types | DiGiammarino et al., *Nature Structural Biology* 9:12–16 (2002), doi:10.1038/nsb730, demonstrated **pH-dependent destabilization of the mutant p53 tetramer**. | A change to any particular TP53 **heterodimer**, such as TP53–MDM2 (a separate tracked pair). |
| **STAT3 Tyr640Phe** | SH2 domain; STAT3–STAT3 is in the union in 154/154 cell types | Mori et al., *Scientific Reports* 7:15919 (2017), doi:10.1038/s41598-017-15912-6, reported **stabilization of activated STAT3 homodimers**. Pilati et al., *Journal of Experimental Medicine* 208:1359–1366 (2011), doi:10.1084/jem.20110283, also investigated activating STAT3 mutations in hepatocellular adenoma. | A direction or magnitude for STAT3–STAT1 or any other specific heterodimer, or condensate occupancy. |

Testing mutations' effects of the PPI cascade performance at the heterodimer level would require a set of specific experiments for each source variant to provide the missing link of the experimental allele-specific evidence for an identified heterodimerization partner. A well-reasoned test would select an interaction whose source mutation lands in a mapped partner-contact region; compare WT and mutant proteins at matched abundance by a direct binding or interaction assay (purified reconstitution or quantitative co-IP/BRET/FRET) in both relevant cell contexts; and then test pair localization and material-state biological behavior separately. Loss of binding, gain of binding, or altered stoichiometry each should be evaluated and considered. Thus, the broader list of mutations should be regarded as a list of candidates for prioritized experimental screens, not a list of established protein dimerization or cascade disruption events.

**DrLLPS and PhaSePro evidence-tiered provenance of the PPI cascade constituents' interactions with structural-functional elements of the liquid-liquid phase-separated (LLPS) condensates.**

To enhance the provenance, in addition to 3,941 putative condensate-forming proteins, it was of interest to perform the analysis of the PPI cascade constituents employing the curated DrLLPS and PhaSePro databases that contain the selected evidence-enhanced annotations and carry-out the analyses in the contexts of the HPA of 154 human cell types. The bulk JSON of the PhaSePro database (Mészáros et al., 2020) contains 121 entries, 59 human, curated as experimentally supported LLPS drivers; 56 human entries mark in vivo and 50 in vitro evidence. The DrLLPS database file (Ning et al., 2020) has 9,281 data rows in the version fetched, among which there are 3,633 human records, comprising 86 scaffolds, 357 regulators and 3,190 potential clients. A DrLLPS regulator can alter phase separation without itself driving it; a client may be associated by interaction or proteomics and is not equivalent to a demonstrated driver. Entries were mapped by UniProt accession (isoform suffix removed) to current reviewed human primary genes, without guessing on unresolved aliases; 3,604 DrLLPS human genes and all 59 PhaSePro human genes were mapped.

Among 1,122 network symbols (1,011 current HPA/UniProt primary human symbols), 13 map to PhaSePro human drivers, 18 to DrLLPS human scaffolds, 37 to DrLLPS regulators, and 166 to DrLLPS clients (Figure 5). These role totals overlap: 12 of 13 PhaSePro network drivers have DrLLPS entries, mostly scaffolds. There are some notable disagreement: ESR1 is a PhaSePro driver but a DrLLPS regulator, and GATA3 has no DrLLPS record. Here, the evidence-tier precedence used in the per-gene table is PhaSePro driver > DrLLPS scaffold > regulator > client > no record. Ten of the 13 (77%) PPI cascade network PhaSePro drivers and 12 of 18 (67%) PPI cascade network DrLLPS scaffolds have a pathogenic feature-overlap mutation annotation. Of the 13 PhaSePro PPI cascade network drivers, APP, DLG4, EWSR1, HNRNPA1, NPM1, PML, SQSTM1, TARDBP, ESR1 and GATA3 have that annotation; GRB2, NCK1 and POLR2A do not. At nTPM ≥1, median per-cell branch plateau counts, there are 10 PhaSePro drivers and 15 DrLLPS scaffolds for TF-HUB PPI cascade branch, versus 3 and 4 for TF–TF PPI cascade branch. These counts concern wild-type gene reach, not mutant protein in HPA cells.

Among genes that have any pathogenic mutation recorded, the feature-overlap proportion is modestly higher in PhaSePro drivers (42/47 versus 1,699/2,320 in other mutated candidates; OR 3.07, BH q = 0.035 across four evidence-tier tests) and DrLLPS scaffolds (53/61 versus 1,688/2,306; OR 2.43, q = 0.035). DrLLPS regulator and client contrasts do not pass BH correction. These signals of statistical significance disappear in a model adjusting for total catalogued mutation burden, protein length, MID annotation and all four overlapping evidence flags.

The global protein dimers test across 154 human cell types was conducted on the 51,439 reconstructed heterodimer types spanning 5,919,637 cell–pair occurrences, now annotating endpoints by four distinct LLPS evidence roles rather than Banani et al. (2022) condensate compartments. In each of 100,000 permutations, all four role labels were shuffled jointly within bins of retained PPI-graph degree and HPA across-cell detection breadth (Figure 5; Table 8). The pair graph, cell-frequency weights, role overlap and role counts within strata were held fixed; upper-tail permutation p-values were BH-adjusted across four role tests and a secondary test counts each pair type once.

The PhaSePro describes partner-dependent LLPS for several human drivers with matching tracked PPI pairs: DLG4–SYNGAP1 (66 cell types; SynGAP fragment required for the described postsynaptic-density phase transition); GRB2–LAT (26 cell types) and GRB2–SOS1 (153 cell types; together part of a multivalent LAT signalosome); and NCK1–NPHS1 (92 cell types) and NCK1–WASL (151 cell types; Nephrin/NCK1/N-WASP system). DLG4 has the Asp332Gly mutation in a source-mapped PDZ-domain interval (314–394) and is reached at the TF-HUB PPI cascade plateau in 143 cells, a strong nomination for a partner-specific assay candidate to seek the experimental evidence that this variant changes SynGAP binding. APP–PRNP (154 cell types) also matches a PhaSePro partner description, but the curated experiment concerns amyloid-β oligomer/PrPc, which is not necessarily reflect the full-length APP–PRNP. PML–SUMO1 likewise suggests the involvement of a SUMO-associated mechanism, not necessarily directly relevant to a noncovalent protein dimer. Observed curated-driver overlap is not uniform across the ten GO compartments significant in the earlier protein dimers annotation test (Table 5): 373/4,827 transcriptional-condensate-proxy pair types and 36/109 PML-body pair types include a PhaSePro driver; 0/9 Cajal-body and 0/13 DNA-replication-condensate-proxy pairs do.

**Table 8.** The global pair test on the 51,439 reconstructed heterodimer types spanning 5,919,637 cell–pair occurrences annotating endpoints by four distinct LLPS evidence roles.

| Endpoint label on both partners | Human role-labeled union endpoints | Observed cell–pair occurrences | Observed / null mean | BH q | Distinct pair types | Distinct-edge BH q |
|---|---|---|---|---|---|---|
| **PhaSePro curated driver** | 54(59) | 8,328 | 2.89 | 0.000587 | 63 | 0.000693 |
| **DrLLPS scaffold** | 80(86) | 17,197 | 2.51 | 0.000440 | 123 | 0.000693 |
| **DrLLPS regulator** | 290(357) | 61,640 | 2.02 | 0.000240 | 463 | 0.000440 |
| **DrLLPS potential** | 2,453(3,190) | 1,004,795 | 0.974 | 0.632 | 7,823 | 0.657 |

| | |
|---|---|
| **client** | |

Legend: The driver/scaffold results survive distinct-edge weighting, but shared citation, assay and protein-network topology beyond degree bins remain possible explanations. Numbers in parenthesis indicate the total numbers of corresponding endpoints.

Overall, there are significant fractions of proteins defined as components of the LLPS biomolecular condensates among protein dimers of the PPI cascade constituents: 54 of 59 (92%) PhaSePro-curated LLPS condensate drivers; 80 of 86 (93%) of DrLLPS scaffold proteins; 290 of 357 (81%) of DrLLPS proteins defined as LLPS condensate regulators; and 2,453 of 3,190 (77%) of DrLLPS proteins defined as potential clients of LLPS condensates (Table 8). Taken together, reported above evidence appear in agreement with the concept postulating the contributions of the PPI cascade chain reactions initiated at DNA templates encoded by TEs to the biogenesis of protein heterodimers having at least one protein monomer selected from condensate-forming protein molecules.

**Evidence of the PPI cascade family H350 constituents' engagements in synaptic phase separated condensates.**

The reported H350 family of the PPI cascade TF-HUB branch contains 350 individual gene symbols (Glinsky, 2026a; b; c), 245 of which unambiguously match current HPA/UniProt primary human genes. The SynGO 2024 synaptic gene ontology database annotations assembled in the Harmonizome 3.0 SynGO gene–term edge list (last updated 4 March 2024) contains 1,593 genes, 267 terms, 3,906 gene–term associations. A total of 94 of 245 mapped the PPI cascade H350 family genes have at least one SynGO synaptic annotation. More specifically, 23 of 245 genes have an explicit PSD cellular-component assignment in at least one of GO:0014069 (PSD), GO:0099092 (PSD intracellular component), GO:0099031 (anchored PSD membrane), or GO:0099061 (integral PSD membrane).

The cross-database analysis of the evidence ladder (Table 9) identified the 12-gene shortlist comprising of DLG4, FYN (DrLLPS scaffold); GSK3B (regulator); and ARF1, ARRB1, CAMK2A, CTNNB1, GRIN1, HSPA8, MAPK1, PAK1, RPS3 (potential clients). DLG4 is the only one of these 12 with a PhaSePro entry explicitly assigned to the PSD. Assigned to the FYN gene the DrLLPS scaffold role and the PSD tag coexist, but they are not by themselves proof that FYN independently scaffolds a PSD condensate. Likewise, DrLLPS tags on HNRNPA1/XPO1 outside this strict SynGO-PSD shortlist was not upgraded to PSD-specific driver evidence, in part, because the PhaSePro assigns HNRNPA1 to stress granules, not to the PSD.

**Table 9.** Cross-database evidence ladder of the PPI cascade family proteins' engagements in post-synaptic density (PSD) phase-separated condensates.

| Evidence level among H350 family proteins | Count | Meaning |
|---|---|---|
| **SynGO synaptic genes, any term** | 94 | Expert-curated synaptic association; not an LLPS annotation. |
| **SynGO explicit PSD cell content term** | 23 | PSD location or membrane subcomponent. |
| **DrLLPS records with PSD tag** | 74 | 63 clients, 7 regulators, 4 scaffolds; **not** 74 experimentally established PSD drivers. |
| **Both any SynGO term and DrLLPS PSD tag** | 45 | Synaptic association plus a condensate-database PSD label; sources may share evidence. |
| **Both explicit SynGO PSD cell content and DrLLPS PSD tag** | **12** | Highest-precision location × database shortlist; individual role strength differs. |
| **PhaSePro experimentally curated PSD driver** | **1: DLG4** | PhaSePro documents PSD-95/SynGAP partner-dependent phase separation. |

These groups overlap and are not additive and assignments were reviewed manually on a case by case basis. Overall, a DrLLPS PSD tag is broad: 1,380 human source records carry it, including 1,313 clients. Among the 74 tagged H350 hub proteins (Table 9), 29 have no SynGO entry in this particular 2024-derived gene–term database table; however, no entry in the 2024-derived database should not be considered as a proof of absence from synapses. Conversely, 11 H350 genes have explicit SynGO PSD annotation without a DrLLPS PSD tag.

Available direct experimental evidence provide strong mechanistic anchor to the presented herein theoretical framework. The phase transition of a PSD-95/SynGAP system has been demonstrated experimentally (Zeng et al., 2016). A multicomponent PSD scaffold system including PSD-95, GKAP/DLGAP1, Shank3 and Homer3 has been successfully reconstituted (Zeng et al., 2018). DLG4 is the PPI cascade family H350 protein (Glinsky, 2026a; b; c). SYNGAP1 and DLGAP1 are not in the reported list of 1,122-symbol PPI cascade constituents, although the identified protein dimers supplied by the PPI cascade contains DLG4–SYNGAP1 (66/154 human cell types) and DLG4–DLGAP1 (142/154 human cell types). SHANK3 and HOMER3 are known DLG4 partners of a multicomponent PSD scaffold system (Zeng et al., 2018). However, they have no listed DLG4 PPI partnership; that is an input-library limitation, not evidence against their established multicomponent assembly.

Among the strict shortlist of the H350 family proteins, the PPI-defined dimers also contain DLG4–FYN (141/154 human cell types), DLG4–GRIN1 (41/154 human cell types) and DLG4–CAMK2A (40/154 human cell types); these three dimers are PPIs' tracked, not individually demonstrated LLPS-driving dimers. Among mutations potentially disrupting the condensate-promoting features, Banani et al (2022) lists DLG4 Asp332Gly within a PDZ-domain interval (314–394). Collectively, these observations make the PSD-95/SynGAP dimer readily amenable for direct experimental examination of whether this allele changes protein-protein binding or condensate assembly and/or functions.

The Figure 6 highlights DLG4/PSD-95–SYNGAP1 as the strongest partner-resolved candidate and DLG4–DLGAP1 as a route supported by multicomponent PSD reconstitution. It also shows the 12-hub proteins as evidence-supported shortlist and maps DLG4 Asp332Gly to a PDZ interval. The variant's effect on binding or phase-separated condensation remain to be experimentally tested. The PPI cascade H350 family contains a large number of proteins annotated to the synaptic gene ontology terms, including DLG4/PSD-95, an experimentally curated postsynaptic-density (PSD) phase-separation driver. The intersection also includes other PSD-localized the PPI cascade H350 family proteins carrying DrLLPS PSD labels, but most of those are potential clients or regulators, not experimentally demonstrated PSD condensate drivers. The strongest evidence-supported protein-protein interaction's partner-resolved route is the DLG4–SYNGAP1 dimer; its presence is inferred from the cell-specific PPI eligibility analyses in the 66 of 154 human cell types.

**Condensate-negative proteins of disordered and structured protein classes are absent among the PPI cascade constituents.**

Pintado-Grima et al (2025) described the panel of 2,876 proteins defined as a collection of comprehensive datasets for liquid–liquid phase separation studies. In the datasets, there are 755 condensate-positive proteins designated as the exclusive driver (DE; n = 337), the exclusive client (CE; n = 363), the client-and-driver (C_D; n = 55); and the control panel of 2,121 condensate-negative proteins (structured condensate-negative protein, NP; n = 1,120; disordered condensate-negative protein, ND; n = 1,001). The released datasets span multiple species and matching the release accessions exactly to a current reviewed human UniProt accession-to-primary-gene table yields 495 human proteins: 382 condensate-positive proteins and 113 condensate-negative proteins (14 NP and 99 ND). This mapping was deliberately conservative and it does not infer species from symbols, project orthologs, or force archived accessions to current genes. The human accession retrieval was downloaded from UniProt (organism_id:9606) and (reviewed:true) on 14 October 2026. All 495 human proteins are mapped to distinct primary symbols. The panel of 1,011 PPI constituents contain 67 of 382 (17.5%) condensate-positive proteins, zero condensate-negative protein of the structured condensate-negative proteins (NP) and zero condensate-negative disordered proteins (ND). For condensate-positive versus condensate-negative proteins, the 2 × 2 contingency table is [[67, 0], [315, 113]], with rows mapped panel/outside and columns positive/negative. Two-sided Fisher exact $p = 8.80 \times 10^{-9}$; the raw sample odds ratio is infinite because no negative is in the PPI cascade panel. With the 0.5 value imputed only to the zero cell for the inference of a finite descriptive estimate of odds rations, OR = 48.1 (approximate log-Wald 95% CI 2.95–783). The difference in observed panel fractions is 17.54 percentage points. The exact binomial 95% upper bound on negative panel membership is 3.21% (implying not zero risk). The exploratory decomposition of the pooled condensate-negative proteins into NP only and ND only sets generated the following outcomes. For the NP panel alone, two-sided Fisher exact test p = 0.140 (only 14 matched human proteins; its zero-count exact 95% upper fraction is 23.16%). For the ND panel alone, Fisher exact test $p = 1.02 \times 10^{-7}$ (99 matched human proteins; upper fraction 3.66%). Overall, the PPI cascade network constituents contain 6.6% of condensate-positive proteins, while the fractions of condensate-positive proteins are 3.7% (n = 26); 4.0% (n = 22); 2.4% (n = 4); 11.4% (n = 28); and 16.7% (n = 41) among the initiator TFs; principal initiator TFs; complementary initiator TFs; TF-TF family PPI branch constituents; and TF-HUB protein family PPI cascade branch constituents, respectfully. Next, the reconstructed protein heterodimer graphs (availability threshold nTPM ≥ 1) were interrogated, comprising of 51,439 distinct non-self-protein dimer types, 5,919,637 cell–pair occurrences across 154 cell types, and 8,873 endpoints (Table 10).

For each of 20,000 permutations, complete cascade-positive/DE/CE/C_D/D+ label vectors are shuffled jointly within reconstructed-graph degree quintile × HPA detection-breadth quintile. Nulls preserve label counts and co-occurrence within strata, the graph and its cell-frequency weights. The statistical endpoints were the upper-tail empirical p = (exceedances + 1)/20,001; and BH correction over the five pre-specified labels, separately for cell-frequency-weighted and distinct-edge statistics (Table 10).

**Table 10.** Summary of the statistical significance evaluations of the interrogations of the reconstructed protein heterodimer graphs across the 154 human cell types.

| Both endpoints annotated | Labeled graph endpoints | Cell–pair occurrences, observed / null | Weighted BH q | Distinct edge types, observed / null | Edge BH q |
|---|---|---|---|---|---|
| **Any positive** | 320 (382) | 125,447 / 44,532 (2.82×) | 0.000125 | 998 / 358.1 (2.79×) | 0.000125 |
| **DE** | 129 (146) | 21,543 / 8,211 (2.62×) | 0.000417 | 171 / 65.8 (2.60×) | 0.000167 |
| **CE** | 157 (199) | 16,228 / 8,015 (2.02×) | 0.0114 | 125 / 66.5 (1.88×) | 0.0168 |
| **C_D** | 34 (37) | 5,895 / 884 (6.67×) | 0.000125 | 47 / 6.56 (7.16×) | 0.000125 |
| **D+** | 48 (50) | 5,667 / 1,393 (4.07×) | 0.000562 | 43 / 10.9 (3.94×) | 0.000562 |

Legend: Numbers in parenthesis are the numbers of all human proteins assigned to the corresponding endpoints.

Using all 20,190 reviewed-human primary gene symbols as the reference, 8,822 (43.7%) proteins are endpoints in the cell-type-reconstructed non-homodimers graph. The Table 11 reports the results of tests whether each annotation overlaps those endpoints more than expected from that reference, without restricting the background by measured expression or PPI-library coverage.

**Table 11.** Overlaps of all five classifications of condensate-positive proteins with PPI cascade constituents are significant under the genome-wide-human reference after correction for five tests.

| Annotation | Graph endpoints / identified humans | Expected endpoints | Observed / expected | Odds ratio (exact 95% CI) | Two-sided Fisher p | Holm-adjusted p |
|---|---|---|---|---|---|---|
| **Any positive** | 320 (382) | 166.9 | 1.92× | 6.86 (5.21–9.17) | $7.81 \times 10^{-60}$ | $\mathbf{3.91 \times 10^{-59}}$ |
| **DE** | 129 (146) | 63.8 | 2.02× | 9.91 (5.95–17.55) | $1.32 \times 10^{-29}$ | $\mathbf{5.29 \times 10^{-29}}$ |
| **CE** | 157 (199) | 87.0 | 1.81× | 4.89 (3.45–7.05) | $3.20 \times 10^{-24}$ | $\mathbf{9.59 \times 10^{-24}}$ |
| **C_D** | 34 (37) | 16.2 | 2.10× | 14.66 (4.61–74.60) | $1.43 \times 10^{-9}$ | $\mathbf{1.43 \times 10^{-9}}$ |
| **D+** | 48 (50) | 21.8 | 2.20× | 31.09 (8.15–264.12) | $2.06 \times 10^{-15}$ | $\mathbf{4.12 \times 10^{-15}}$ |

Legend: Numbers in parenthesis are the numbers of all human proteins assigned to the corresponding graph endpoints. D+ classification overlaps refer to the other condensate-positive annotations, so the rows are not independent categories.

In summary, the results of the above analyses demonstrate that there are no condensate-negative proteins of either structured protein category or disordered protein category among the PPI cascade constituents. In contrast, protein molecules assigned to all analyzed categories of condensate-positive proteins are significantly overrepresented among the PPI cascade constituents. In total, 320 of 382 (84%) of all condensate-positive proteins are among the PPI cascade constituents (Table 11). Consistently, 129 of 146 (88%) of DE proteins; 157 of 199 (79%) of CE proteins; 34 of 37 (92%) of C_D proteins; and 48 of 50 (96%) of D+ proteins constituting various condensate-positive protein categories are significantly overrepresented among the PPI constituents' network of protein dimers (Table 11).

**AlphaFold 3D modelling of mutations' patterns and analysis of deposited partner-bound experimental structures of a measured protein dimerization interface.**

To better understand the potential overall impacts of human disease-associated mutations with phase-separated condensate-disrupting potentials on protein dimerization, the mutation positions were mapped onto AlphaFold DB wild-type structures for the 299 PPI cascade constituents. This analysis is the target amino acid residue mapping on the wild-type 3D structures, not de novo modeling of mutant proteins. It was applied to 299 PPI cascade constituents harboring any human disease-associated pathogenic mutations, including 269 PPI cascade constituents that have a pathogenic mutation in genes encoding condensate-forming proteins (Figure 4).

At the time of the analysis (10/05/2026), AlphaFold DB models were available for 297 of 299 genes (AlphaFold DB run 2026-10-05). Across them, 17,947 variant-label records passed the sequence and coordinate checks, representing 12,861 distinct accession mapped–residue sites (Table 10). AlphaFold DB had no models for APC or HTT proteins. Gene symbols reflect the UniProt accessions; since the CUX1 symbol has two source accessions, there are 300 protein-accession records for 299 genes (Table 12). A total count of 299 genes provides all catalogued human pathogenic mutation labels, while a set of 269 genes captures mutation labels corresponding to the putative condensate-promoting features.

**Table 12.** Summary statistic metrics of the 3D modelling using the AlphaFold DB run 2026-10-05.

| Measure | Result |
|---|---|
| **Genes in requested intersection** | 299 |
| **Source accession records** | 300 |
| **AlphaFold DB wild-type models retrieved** | 298 records / 297 genes |
| **Genes with ≥1 sequence-validated site** | 295 |
| **Unique mapped accession–residue sites** | 12,861 |
| **Pathogenic-label records parsed from S4C/S4H*** | 22,195 |
| **Successfully mapped records** | 17,947 |
| **S4H feature-overlap records mapped*** | 9,038 of 12,237 |
| **S4H missense records mapped*** | 4,493 |
| **Mapped H labels at local pLDDT < 70** | 4,266 of 9,038 |

Legend: *, human pathogenic mutation label records were parsed from the Supplementary Table S4C and S4H reported in the Banani et al. (2022). Each label is checked against the 2020_06 source sequence, current AlphaFold DB reference sequence, model fragment boundaries, and the PDB Cα residue identity. Only labels passing every check get coordinates and pLDDT. For missense and single-residue deletion/duplication/stop labels, the coordinate is the wild-type residue. For frameshifts, it is the start of the frameshift, not the altered downstream chain. Range indels, unknown starts, extensions and other non-single-site notations are not given an arbitrary single coordinate. Duplicate variants at one site remain separate records in the CSV; colored markers represent unique sites, not events. APC (P25054; 2,843 amino acids, 2,221 mutation labels) and HTT (P42858; 3,142 amino acids, two mutation labels) have no AlphaFold DB prediction at these accessions. They are not presented as modeled. RFX7 and SRRM2 have an available model but none of their source labels passes the site/fragment/reference checks; their wild-type 3D models are in the archive but there is no mapped colored variant site.

Among all 22,195 label records, 1,202 have no unique single-residue coordinate, 663 lie outside the retrieved model fragment, 158 fail the current-model reference residue check, and two fail the source reference sequence check. The other 2,223 belong to the two un-modeled proteins APC and HTT. Such records remain in the Supplementary table variant_coordinates.csv with an explicit status rather than being positioned by guesswork. The 663 fragment exclusions are dominated by CHD7. The 158 sequence disagreements are dominated by CHD7 and MYC; inspect isoforms/sequence-version changes before using those labels in structural claims. AlphaFold DB 3D models with clearly labelled mutation residue positions are available for the online visualization in the Supplementary Information as searchable 3D viewer for all 299 genes designated as variant_structure_viewer.

AlphaFold DB v6 wild-type monomer models cannot tell whether a mutation changes binding, condensate formation, or PPI cascade dynamics. Next, tor DLG4; CTNNB1; HSPA8 (Figure 7), the binding partners and 3D model protein dimers were selected highlighting mutations relevant to the binding interface. DLG4–SYNGAP1, CTNNB1–TCF7L2, and HSPA8–BAG1 dimers were selected because each dimer candidate-pair census and has an experimentally resolved partner-bound structure (Table 13). The 3D viewer highlights the interfaces and maps the human disease-associated mutation labels onto them displaying experimentally-validated bound fragments, not AlphaFold-Multimer predictions.

**Table 13.** Human disease-associated mutations relevant to the experimentally-validated binding interface of DLG4–SYNGAP1, CTNNB1–TCF7L2, and HSPA8–BAG1 dimers.

| Target–partner | Experimental structure | Structural ascertainment caveats | Candidate-pair cell types |
|---|---|---|---|
| **DLG4–SYNGAP1** | 5JXB, X-ray, 2.9 Å | Human DLG4 PDZ3 bound to a mouse SYNGAP C-terminal motif; crystallized as an engineered single-chain fusion, split into two display chains. Not a free human–human dimer. | 66/154 |
| **CTNNB1–TCF7L2** | 1JDH, X-ray, 1.9 Å | Separate bound human fragments: CTNNB1 armadillo region and a TCF7L2 peptide. | 153/154 |
| **HSPA8–BAG1** | 3LDQ, X-ray, 1.9 Å | Separate human HSPA8 ATPase and BAG1 BAG-domain fragments; structure also contains a small molecule, excluded from protein–protein distance measurements. | 154/154 |

The human disease-associated mutations variant catalog of Banani et al. (2022) lists genes with any disease-associated labels and genes with mutations located within candidate condensate-promoting features. Mutations are displayed at the reference residue only when the original label and PDB residue agree. Interface contact is called when a minimum distance between protein heavy atoms ≤5 Å; at distance of 5–10 Å it is called nearby, not directly contacting (Figure 7). 3D viewer of human disease-associated mutations vis-à-vis the interface for 3 protein dimers is available online in the Supplementary Information as the file designated three_partner_complexes. The clearest interface overlap is CTNNB1–TCF7L2: human disease-associated mutations linked to variants potentially disrupting the condensate-promoting features occur at three directly contacting CTNNB1 sites, Lys292, Lys335, and Asn387. DLG4 Asp332Gly is 8.09 Å from the SYNGAP motif, which places the mutation nearby, not in a direct contact zone. None of HSPA8 protein's three sites directly contacts BAG1 in the experimentally-resolved fragment (Figure 7). It should be noted that the proximity alone does not establish a mutation's effect (or prove the absence of the effect) on protein-protein binding.

Overall, the main findings (Figure 7) and limitations for human disease-associated mutations potentially affecting condensate-promoting features of protein dimers of the PPI cascade constituents could be summarized as follows:

- DLG4–SYNGAP1 dimer: DLG4 Asp332Gly is 8.09 Å from the motif (near, not in direct contact). DLG4 Gly198Ser, Val476Gly and the Tyr50Leufs starting site lie outside the resolved PDZ3 construct. Separately, PDB 6QJK is an experimentally determined isolated DLG4 D332G PDZ3 mutant domain, included for comparison; it is not a mutant SYNGAP-bound complex and cannot establish altered binding.
- CTNNB1–TCF7L2 dimer: Three distinct human disease-associated mutation reference sites touch TCF7L2 in this bound fragment: Lys292 (3.47 Å; three different amino-acid substitutions), Lys335 (2.89 Å; two substitutions) and Asn387 (3.75 Å; four substitutions). These nine labels mark three sites; they are not likely to represent nine independently observed interface-affecting mechanisms. Several other mutation sites lie 5–10 Å away, while the densely labelled N-terminal degradation region is outside of this structure. Mutation proximity to the contact site suggests, but not establish, that specific amino acid residue replacement weakens or strengthens TCF7L2 binding.
- HSPA8–BAG1 dimers: No mutation-labelled residue directly contacts BAG1 in this structure. Ala133Ser is resolved but 22.96 Å from the contact with the partner; Thr429Ser and Gln473Arg lie outside the crystallized ATPase fragment. This complex cannot answer whether the latter two mutations affect a substrate-binding-domain interaction with a different protein partner.

The three examined herein protein dimers have respectively 4, 109, and 3 labels of human disease-associated mutations potentially affecting the condensate-promoting features. Importantly, the 5 Å contact cutoff measures the interface geometry in the deposited wild-type structure, not binding energy or an effect of a mutation on protein-protein binding. Molecular assemblies, oligomerization state, stoichiometry, cell-specific co-localization and phase separation are not established by these structures or by the candidate PPI census. All these features remain to be experimentally evaluated to determine the precise role of specific human disease-associated mutations potentially affecting the condensate-promoting features in pathogenesis of human disorders.

**Discussion**

Recent observations summarized in the Introduction provide a theoretical foundation supporting the concept of Hybrid Assembly Pathways (HAP) of endogenous multiprotein complexes (EMCs). The cornerstone idea at the origin of the HAP model is the concept that ~4.5 million TE-derived genomic loci predominantly residing in transcriptionally-silent chromatin regions of differentiated human cells function as the nucleation sites for the initiation of protein-protein interactions (PPI) cascades for creation of cell type-specific pools of protein homo- and heterodimers. The PPI cascades operate concomitantly and iteratively on arrays of DNA templates to continually create protein dimers, which upon the release from DNA templates serve as attractors for on-demand recruitment of protein monomers to finalize the assembly of EMCs of desired compositions. Initial sets of evidence at the inception of the HAP concept were obtained from the analyses of transcription factor binding sites (TFBS) encoded by TE-derived DNA sequences and interrogations of TFs bound to genomic loci that are engaged in execution of divergent species-specific regulatory programs of primates' brain development, including Modern Humans, Chimpanzee, and Rhesus Macaque (Glinsky, 2026a).

Subsequent explorations of the Modern Human and Chimpanzee brain development genomic regulatory programs revealed the key contributions of PPI cascade chain reactions originated at TE-encoded DNA sequences by a set of 716 proteins defined as the initiator TFs (Glinsky, 2026b). Investigations of distinct panels of DNA-bound initiator TFs and proteins of PPI cascade ensembles demonstrated that they are enriched in either defined sets of neuroanatomical structures (TF-TF PPI cascade mode) or among structural-functional constituents of synapses (PPI cascade mode of TF-HUB proteins). Follow-up analyses demonstrated that the PPI cascades appear to function as biologically-congruent two-branch bifurcated networks of TF–TF constituents that map to transcriptional signatures of hundreds of brain regions, and TF-HUB constituents that map to synaptogenesis and synaptic structures, suggesting the unified logic of genomic functions coordinating region identity and connectivity (Glinsky, 2026b). Most recently, molecular features and inferred biological functions of PPI cascade constituents were elucidated within the context of cell type-specific intracellular protein monomer pools documented in the Human Protein Atlas (HPA) of 154 human cell types (Glinsky, 2026c). It has been demonstrated that PPI cascade constituents have the capacity to generate 51,830 protein dimers within the proteome availability contexts of 154 types of human cells, which is consistent with the proposed role of PPI cascade constituents in biogenesis of structured multiprotein complexes in human cells. Further, potentially important pathological impacts were revealed of the persistence of proteinopathy proteins by causing the damage to the PPI cascade performance by sequestering 44 synaptic proteins (Glinsky, 2026c).

In this contribution, multiple lines of analytical evidence has been systematically acquired and interrogated to test the validity and potential practical utility of the HAP concept to shed the light on the mechanisms of biogenesis of liquid-liquid phase-separated membraneless biomolecular condensates. Analytical evidence reported in this contribution demonstrate that condensate-forming proteins are significantly enriched among the PPI cascade constituents and support the concept of contributions of PPI cascade constituents to the biogenesis of intracellular pools of cell type-specific protein heterodimers assembled with the engagement of at least one protein monomer selected from condensate-forming protein molecules. It has been documented that there are no condensate-negative proteins of either structured protein category or disordered protein category among the PPI cascade constituents. In contrast, protein molecules assigned to all analyzed categories of condensate-positive proteins are significantly overrepresented among the PPI cascade constituents.

One of the useful approaches for identification of high-priority protein heterodimers for follow-up structural-functional studies of potential fundamental and translational significance is the mapping of protein monomers encoded by genes harboring human pathogenic mutations. There are 409 unique mapped genes encoding putative condensate-forming proteins among the 1,011 PPI cascade constituents; a total of 299 of those (73%) have at least one human pathogenic mutation and 269 (90%) have a human condensate-disrupting pathogenic mutation potentially affecting condensate-forming proteins. It has been noted that large fractions of the PPI cascade-supplied heterodimers have proteins with candidate condensate-disrupting mutations. Of 51,439 reconstructed heterodimer types, 22,546 (43.8%) have ≥1 candidate condensate-disrupting mutations and 3,325 (6.5%) have two. Across 5,919,637 cell–heterodimer occurrences, the corresponding incidences are 2,705,299 (45.7%) and 415,171 (7.01%).

Taking into considerations the essentially unlimited availability of DNA templates, the PPI cascade network has the capacity to run concurrently both the LLPS condensate arm and the structured multiprotein complexes of the CORUM arm with no anticipated competition: both reach a plateau in every cell type, from the same 716 seeds of initiator TFs. Notably, the PPI cascade TF-TF branch's plateau is richer in the condensate-forming proteins and the plateau of PPI cascade TF-HUB branch is richer in the CORUM multiprotein complexes, consistently, in 154 of 154 human cell types. However, the T245 TF-TF PPI cascade and H350 TF-HUB proteins families are not the two functionally-segregated PPI cascade branches, or the separate populations because both of the families manifest clear cross-annotation profiles: 38.8% of the PPI cascade H350 plateau is LLPS condensate-annotated and 57.7% of the PPI cascade T245 plateau is in CORUM multiprotein complexes.

Despite the employment of the comprehensive set of analytical approaches implemented in this contribution, this study has several limitations. The catalogue of protein homo- and heterodimers that could be potentially created by the PPI cascade operations was estimated based on the analysis of availability of proteins expressed in 154 human cell types, which utilized the most comprehensive available to date data set of the Human Protein Atlas. However, in each human cell type the intracellular availability of protein is expected to be highly dynamic and undergo fluctuations in response to intracellular regulatory cues and extracellular inputs. Essentially all components of intracellular dimer pools defined analytically in this contribution and other studies would require the experimental validation. Similarly, the precise mechanistic roles, if any, of human pathogenic mutations on protein dimerization and anticipated structural-functional consequences of their potential condensate-disruptive impacts remain to be determined. As illustrated in this work, applications of the AlphaFold reference models and other approaches to mutation-site validation analyses should be helpful in narrowing the broader panels of candidate protein dimers to high-value list of targets nominated for in-depth structural-functional interrogations in vitro and in vivo. Outcomes of these and many other experimental studies would ultimately determine the realistic availability and effectiveness of emerging therapeutic opportunities of PPI targeting stoichiometric interventions designed to rescue the pathological defects of altered performances of biomolecular condensates.

## Conclusion

This contribution documents the multiple layers of analytical evidence supporting the concept that ~4.5 million TE-derived genomic loci comprising at least half of human genome and predominantly residing in transcriptionally-silent chromatin regions of differentiated human cells function as the nucleation sites for the initiation of protein-protein interactions (PPI) cascades for creation of cell type-specific pools of protein homo- and heterodimers. While the reported analyses were focused on human cells, considering the ubiquitous TE colonization in eukaryotes, the reported observations are likely relevant to any multicellular organisms. In the evolutionary context these observations imply that accumulation of genomic DNA sequences encoded by TEs might be beneficial (perhaps, required) for evolutionary fitness because this feature provides the structural frameworks facilitating the execution of any of the biologically-essential functions in eukaryotic cells, enactment of which necessitates the assembly of structured and membraneless endogenous multiprotein complexes. This contribution identified numerous protein heterodimers that are potentially highly relevant to physiology and pathology of Modern Humans for follow-up mechanistic studies of potential fundamental and translational impacts.

## Methods

### PPI cascade constituents, liquid-liquid phase separated condensate annotations, variant mapping of human pathogenic mutations, structural interfaces, and functional inferences

This section reports all essential analytical and statistical components of the Methods implemented in this contribution. Additional descriptions of methodological, analytical, and technical aspects of this study are available online in the Supplementary Information section of the paper, including the files of the Read Me guides; the Supplementary Summary Files; as well as available upon request archives of supplementary files of condensate project csv tables and codes.

### Study design and units of analysis

The supplied cascade contained 1,122 distinct protein symbols (Glinsky, 2026a; b; c). Its 716 initiator transcription factors (TFs) had binding sites at 600 LTR5_Hs-encoded loci in the human genome, with additional TE-family analyses used to confirm the initiator designation. Subsequent GSEA distinguished 545 principal from 171 complementary initiator TFs. GSEA of the 716 initiators against the TF–TF PPI and PPI HUB-protein libraries identified the T245 and H350 cascade branches, respectively. The four family sets are partially overlapped and were not summed to derive the network size. The downstream analyses used branch-specific reached-protein records and a union of 51,830 candidate protein pairs evaluated across 154 Human Protein Atlas (HPA) cell types. A *protein* was the unit for set-enrichment tests, a unique unordered endpoint pair was the unit for distinct-edge analyses, and a pair–cell occurrence was the unit for cell-frequency-weighted descriptive analyses. Self-pairs were retained only for analyses explicitly concerned with self-association and were excluded from heterodimer pairing tests. Proteins could occur in several sets or both reached branches, and a pair could recur across cell types; these overlaps were retained.

Family and initiator/plateau flags were read from family_membership_1122.csv. For analyses requiring current human identifiers, the symbols were intersected with reviewed-human UniProt *primary* gene names; 1,011/1,122 mapped, including 710/716 of initiator TFs, 544/545 of principal initiator TFs (P545), 166/171 of complementary initiator TFs (C171), all 245 of T245 (TF-TF PPI cascade branch); and 245/350 of H350 (TF-Hub proteins PPI cascade branch). Unmapped symbols were excluded from these denominators, not classified as annotation-negative. Reached proteins were read from cascade_entry_levels.csv separately for each cell type, branch (TF–TF or TF-Hub), and the supplied availability rule (labelled nTPM >= 1 or top 4,000 expressed genes). Within each cell × branch × rule, a protein was counted at most once when inPlateau = 1; the number of such records was checked against the reported plateauSize. Branch plateaus and family sets were not combined into disjoint classes (Glinsky, 2026a; b; c).

**Cell-type expression and conservative pair-incidence reconstruction**

For the cell-resolved pairing analysis, the public HPA version 25.1 rna_single_cell_type.tsv.zip table was matched to all 154 cell-type names in the supplied atlas. The HPA download called its expression column nCPM (normalized counts per million)**;** the analyzed dimer supplement utilize numeric expression values nTPM (normalized transcripts per million)**.** When more than one HPA Ensembl record carried the same gene symbol, the first source record was retained. Missing gene–cell measurements were set to zero in the reconstructed 8,881-endpoint × 154-cell matrix. A candidate pair was called eligible in a cell if both endpoints had HPA expression **≥1** and at least one endpoint was marked as selected by either cascade branch in that cell under the supplied nTPM >= 1 rule. Selection of only one endpoint without confirming expression of the second was rejected because it systematically overcounted pairs.

The reconstructed incidence for each pair was compared with nCellTypes in dimer_census_union.csv. Only pair types for which the reconstructed number of positive cells equalled the supplied count were retained for the formal all-cell compartment analyses (51,630 of 51,830 pair types). The 200 discordant pair types were excluded rather than assigned to uncertain cell types. After exclusion of 191 self-pairs, the formal heterodimer universe comprised 51,439 pair types and 5,919,637 reconstructed pair–cell occurrences**.** Pair identity in brain excitatory neurons and hepatocytes was checked against the two cell types in the supplied annotated-dimer supplement; these exemplar comparisons served as reconstruction checks, not as independent validation in the remaining 152 cells. The original atlas column labelled nCells[nTPM >= 1] agreed with counts above **1** in the public HPA file for matched genes, whereas the pair supplement included partners reported at exactly **1.0**. We therefore used ≥1 for pair reconstruction, documented the discrepancy, and did not infer the original detection code from the column label alone.

The annotated supplement contained 76,294 pair–cell rows from the two exemplars only. For direct branch-labelled descriptive comparisons, we retained rows with type = heterodimer and call = mutual, and kept tftf, hub, and both as three separate labels. bothEngaged in that supplement represented reciprocal cascade-term annotation, not joint expression. Branch-specific pair assignments for the other cells were treated separately from the directly recorded exemplar labels: compatible TF–TF-only, hub-only, and both-library permissions were retained only when their predicted pair count matched the 154-cell marginal and their positive exemplar branch labels agreed. Any pair–cell assignment on which compatible permissions disagreed was coded unknown, not allocated to a preferred branch. Static library permission remained ambiguous for many pairs; the original branch-labelled graph is needed for definitive assignments.

**Condensate candidate proteins and source-defined compartments**

The putative condensate-forming candidate set was obtained from Banani et al. (2022), Table S4E. It comprises 3,941 UniProt/SwissProt v2020_06 protein entries selected as the **union** of DeepPhase score ≥0.9, the top decile of PSAP sequence scores, and proteins curated in LLPSDB, PhaSePro, or PhaSepDB v1.3. A candidate could enter through any one source. This definition includes computational predictions and is not an assay of phase separation. Table S4A (Banani et al., 2022) supplied the original UniProt accessions; these were mapped to current reviewed human primary gene symbols, and source entries mapping to the same gene were collapsed for gene-level tests. Three historical source entries lacked a current UniProt record. The candidate set was kept distinct from the 35 compartment annotations in Table S4I (Banani et al., 2022).

Table S4I contained 39 GO-term rows grouped by the source into 35 named compartments. We split each source protein list into UniProt entries, resolved accessions through Table S4A (Banani et al., 2022) , mapped them to current reviewed human primary gene symbols, and deduplicated repeated compartment–protein incidences, including repeats across GO terms under the same compartment. A protein could belong to several compartments. The full sheet-I annotation set, including proteins outside Table S4E (Banani et al., 2022), was used as the compartment source. Three asterisk-marked compartment names in the source—DNA repair focus, DNA replication condensate, and transcriptional condensate—used related **functional GO terms** **as proxies,** not direct localization annotations. All other source-defined compartments were also treated as annotations rather than observed phase-separated assemblies.

### Family and reached-plateau annotation analyses

For the static constituent-set screen, each of P545, C171, T245, and H350 was compared with the other network proteins eligible for reviewed-human-gene mapping. This background comprised 1,011 of the 1,122 supplied network symbols; 111 unresolved symbols were not silently counted as annotation-negative, including 105 members of H350 PPI cascade branch. Each family × compartment comparison used a one-sided Fisher exact test for overrepresentation. A further two-sided comparison used proteins exclusive to T245 versus proteins exclusive to H350, excluding shared members. Benjamini–Hochberg (BH) adjusted values were calculated over the combined 175 comparisons (four family-versus-remainder screens and 35 T245-only/H350-only contrasts). These tests concerned fixed family annotation content and did not test reached-protein composition, degree-matched effects, or physical compartment assembly.

For each of the 154 cell types, the observed number of reached proteins carrying each compartment annotation was counted separately in each branch under each availability rule. To ask whether that count exceeded what would be expected from the branch's own repertoire of commonly reached proteins, a null was formed within each branch × rule union of reached proteins**.** Each protein's reach breadth was the number of cell types in whose plateau it appeared. Proteins were assigned to nine breadth strata (1–9, 10–24, 25–49, 50–74, 75–99, 100–124, 125–144, 145–153, or 154 cells). Within each stratum, the compartment-label count was held fixed; for a given cell, the null count was the convolution of stratum-specific hypergeometric draws. One-sided upper-tail exact probabilities were BH adjusted over 154 cells × two branches × two rules × 35 labels (21,560 tests). This within-branch null did not test the difference between TF–TF and hub branches. Compartment fractions and between-branch contrasts were otherwise descriptive, especially for the transcriptional-regulation GO proxy whose definition was related to the TF–TF source library.

### Global heterodimer co-annotation and matched null

For every retained heterodimer type, both-endpoint membership in each of the 35 compartment annotations was determined. The primary statistic for a compartment summed the observed reconstructed nCellTypes frequency of all pair types whose **both** endpoints carried that annotation. Graph degree was computed on the retained 51,439-heterodimer graph; across-cell detection breadth counted HPA cell types in which each of the 8,881 endpoint symbols had expression ≥1. Endpoints were grouped into ten quantile-based, tie-preserving degree bins crossed with five breadth bins (0–38, 39–77, 78–115, 116–153, and 154 cells), giving 40 non-singleton strata in these data.

In each of **100,000** permutations (random seed 20261003), whole 35-label vectors were shuffled jointly among proteins within the same degree × breadth stratum. This preserved each label's stratum frequency and the within-protein overlap among compartment labels. The endpoint graph, each pair's reconstructed cell-frequency weight, and the cell assignments were held fixed. For each label, an upper-tail Monte Carlo probability was calculated as (1 + number of permuted statistics ≥ observed)/100001; the 35 probabilities were BH adjusted. The procedure was repeated as a sensitivity analysis with every distinct heterodimer counted once rather than weighted by its occurrence across cells. The 154 cells were **not** treated as 154 independent realizations of the same interacting edges. A separate per-cell, degree-bin observed/expected ratio was descriptive and was not assigned a permutation p-value. The null did not match publication or annotation study effort, finer graph topology, or physical complex composition.

### Independent tiers of LLPS evidence

Human records in PhaSePro and DrLLPS were mapped to reviewed primary gene names through UniProt accessions (isoform suffixes removed). PhaSePro's experimentally curated driver designation was kept distinct from DrLLPS scaffold, regulator, and potential-client roles; overlapping assignments were retained as simultaneous flags rather than interpreted as independent experiments. Source partner-dependence and cited studies were recorded separately when relating curated LLPS examples to candidate PPI edges. PhaSePro was already an input to the Banani et al. (2022) Table S4E union, so overlap between those sources was not used as independent corroboration.

The global heterodimer co-annotation statistic and 100,000 joint within-degree × detection-breadth permutations described above were repeated for four role indicators—PhaSePro driver, DrLLPS scaffold, regulator, and potential client—while keeping role overlaps together. Upper-tail probabilities were BH adjusted across four roles. A distinct-edge analysis provided a sensitivity check on the cell-frequency weighting. Neither endpoint role membership nor an interaction edge was taken to show that a particular pair forms a condensate.

### Integrated 2025 LLPS roles and source-defined negative proteins

The full v1.1 datasets.tsv release accompanying Pintado-Grima et al. (2025) was used without applying the paper's separate sequence-homology reduction for predictor benchmarking. The 2,876 accession-unique, cross-species records were parsed as mutually exclusive role classes: exclusive driver (DE), exclusive client (CE), client-and-driver (C_D), structured negative from PDB (NP), or disordered negative from DisProt (ND). The release contained 755 positives and 2,121 negatives; overlapping D+ and C+ confidence flags were retained separately. Published accessions were matched exactly against the reviewed human UniProt primary-accession/primary-gene table downloaded on 14 October 2026 with query organism_id:9606 AND reviewed:true. No ortholog projection, historical-alias substitution or inference of human identity from gene symbols was used. The resulting 495 uniquely mapped human records comprised 382 positives, 14 NP and 99 ND. Gene-level membership and rates in the cascade used the 1,011 reviewed-human-primary mapped panel genes, not all 1,122 supplied symbols. A gene absent from the 2025 release was recorded as un-labelled rather than negative.

Association of published positive versus negative status with membership in this independently derived panel was evaluated by two-sided Fisher exact test on the 495 labelled human records. The primary table compared pooled condensate-negative protein sets NP + ND with DE + CE + C_D; separate NP-only and ND-only comparisons were exploratory. For zero cells, the uncorrected sample odds ratio was reported as infinite, while a finite descriptive odds ratio and approximate log-Wald interval were obtained by adding 0.5 only to zero cells. An exact Clopper–Pearson upper bound was calculated for the observed negative-panel fraction. This association was conditional on the source's sampling: the authors had excluded known LLPS entries and first-degree BioGRID interactors of positives from both negative sets. That filter was not a deterministic exclusion from the TE-initiator/GSEA-derived cascade, but it could bias negative representation in a PPI-enrichment-derived panel.

Family role counts were calculated separately for each overlapping family among mapped panel members. Within the 382 mapped human positives, two-sided Fisher comparisons of DE, CE, C_D, D+ and C+ presence in versus outside the panel were BH adjusted across five tests. For an internal connectivity control, complete positive/DE/CE/C_D label vectors were permuted jointly across the 1,011 mapped panel genes within ten deciles of cascade-internal PPI degree, with 20,000 draws (seed 20251014). Family membership stayed fixed, and two-sided empirical probabilities, calculated as twice the smaller of the upper- and lower-tail probabilities (capped at one), were BH adjusted across 20 family × role tests. The H350-only and T245-only portions, after removing their shared genes, were also compared by an exploratory two-sided Fisher test.

To parallel the earlier DrLLPS pairing analysis, the same 51,439 reconstructed heterodimer types and 5,919,637 cell–pair occurrences were annotated by five overlapping indicators: any 2025 positive, DE, CE, C_D and D+. For each of 20,000 permutations (seed 20251014), whole indicator vectors were shuffled jointly within quintiles of retained pair-graph degree crossed with quintiles of HPA nTPM ≥1 detection breadth. Upper-tail probabilities (1 + exceedances)/20001 were BH adjusted over five indicators, separately for cell-frequency-weighted and distinct-edge counts. This whole-graph analysis included recruited endpoints outside the 1,122-symbol panel. A second analysis retained only edges whose two endpoints were among the 1,011 mapped panel genes; its label vectors were shuffled over all 1,011 mapped genes, including those with no retained edge, with the same 20,000-draw stratification (seed 20251015). The graph, cell-frequency weights and within-protein overlap among labels were fixed. Co-annotated endpoints were not interpreted as experimentally confirmed phase-separating pairs.

**Pathogenic-mutation ascertainment and feature overlap**

Primary source of the pathogenic mutations reported in Banani et al. (2022) Supplementary Table S4. Table S4C catalogued all source-listed human pathogenic variant labels, while S4H selected human pathogenic mutation labels computationally assigned to condensate-promoting sequence features in candidate condensate-forming proteins. Source S4A entry names (Banani et al., 2022) and accessions were joined to reviewed current human symbols and collapsed at the gene level, without pooling distinct variant labels across genes. A gene was mutation-ascertained if it had at least one S4C or S4H label, and S4H-positive if any S4H label was present. This allowed the four S4H-positive source entries with empty S4C mutation fields to remain in the annotation analysis. Network/family status and branch-reach breadth were then joined by gene symbol. Disease-label strings supplied in the source were retained at the protein level and were not assigned to individual variants.

For each of the four potentially overlapping families, and for the full mapped network-candidate subset, a two-sided Fisher exact test compared the fraction S4H-positive among mutation-ascertained S4E candidates in the tested set with other mutation-ascertained S4E candidates. BH correction was applied to these five contrasts. As an ascertainment sensitivity analysis, a binomial-logit model for S4H-positive status among mutation-ascertained candidates with a usable S4C total included a network-membership indicator, log protein length, log(1 + number of catalogued S4C mutation labels), and source MID presence. A separate role analysis added four overlapping PhaSePro/DrLLPS role indicators. Genes for which S4H was populated but the S4C total was unavailable were omitted from models requiring mutation burden, rather than assigned zero mutations. These regressions addressed cataloguing differences incompletely and were not interpreted causally.

To screen mapped protein features, single-site missense labels were parsed into reference amino acid and residue number and compared with the MID and low-complexity sequence intervals in the source workbook. Frameshifts, truncations, and range indels were not assigned a missense-specific functional effect by overlap alone. This comparison was a positional annotation, not an enrichment test, because S4H was constructed to include feature-overlapping labels. S4H-positive protein endpoints were overlaid descriptively on reached plateaus and reconstructed heterodimers; these overlays did not imply that a reference HPA cell carried a mutation.

**AlphaFold reference models and mutation-site validation**

The 299 distinct network genes that occurred among genes reported in the Supplementary Table S4E (Banani et al., 2022) and had an S4C or S4H pathogenic label were linked by original source entry name to S4A UniProt accessions. Because one gene had two source accessions, 300 accession-specific records were queried against the EMBL–EBI AlphaFold Protein Structure Database. Available AlphaFold DB version 6 wild-type monomer PDB coordinates and model metadata were downloaded; no de novo mutant fold or AlphaFold-Multimer complex was generated. For every source label, an unambiguous single-residue substitution, stop, deletion, duplication, or frameshift start was parsed into a numbered reference residue where possible. The reference amino acid was checked against the source S4A sequence, the current AlphaFold reference sequence, and the PDB residue identity; positions outside the available model fragment, with missing Cα atoms, or with sequence disagreement were retained with an explicit unmapped status. Single-residue coordinates and local model pLDDT were recorded from the **wild-type** Cα atom. Frameshifts were placed only at their first affected residue; range indels and labels without a unique residue were not assigned a point coordinate. Multiple substitutions at the same reference site remained separate variant labels but one unique structural site.

Two accession records, APC and HTT, had no model available from AlphaFold DB in this retrieval. Model absence was not repaired by borrowing an unrelated isoform or by asserting a predicted structure. AlphaFold pLDDT described local confidence of the reference monomer; it was not a probability that a mutation altered binding, nor a confidence score for an interprotein complex.

### Mutation-label burden and sequence/structure proximity

All 299 PPI cascade network genes overlapping with Banani et al. (2022) S4E data set with at least one S4C or S4H pathogenic variant label were analyzed, including the two without an AlphaFold DB model and genes with too few mappable missense sites for a clustering test. Four variant scopes were counted separately per gene: C-or-H union, H-only, all missense, and H-missense. A *label* was a distinct source variant description; C/H label burden was divided by the source reference length and reported per 1,000 residues. For position analyses, only records whose single-residue coordinate, source reference and current model reference passed the mapping checks above were used. Distinct mapped residue positions, rather than label records, were counted once per gene for each positional statistic. The CUX1 records came from two source accessions of different lengths (678 and 1,505 aa); their labels were retained, but no combined length-normalized burden was assigned.

The label-density distribution among the 294 modelled, position-mapped genes with a single comparable source length provided within-catalogue percentiles. Provisional burden tiers were defined as top decile (≥90th percentile), upper quartile (75th–<90th), and lower/middle (<75th); these were not human-proteome reference rates. The primary one-dimensional hotspot statistic was the largest count of distinct coordinate-mapped **missense** sites in any sliding window of width ceil (0.10 × source length) residues. Every random draw rescanned all windows. For each gene with ≥3 distinct mapped missense sites (175 genes), 5,000 sets of the same number of positions were drawn without replacement uniformly across its source length (random seed 20261016); upper-tail Monte Carlo probabilities (1 + exceedances)/5001 were BH adjusted across the 175 genes. The nine visually selected examples had a separate 20,000-draw, four-scope exploratory analysis; cohort classes used the 175-gene adjusted missense statistic, not those selected-example q-values. Failure to reject the uniform-position null was recorded as *no hotspot detected*, not as evidence of spatial uniformity. Genes with <3 mapped missense sites remained untestable.

For the structural proximity analysis, one wild-type AlphaFold DB accession/model was used for each gene with a suitable model. Distinct coordinate-mapped missense sites at positions with pLDDT ≥70 were retained.

The test statistic was the fraction of site pairs separated by at least 20 residues in sequence whose Cα atoms were ≤10 Å apart. The observed sequence-position pattern was shifted circularly by a nonzero number of residues over the same source sequence; shifts preserved the cyclic pattern of site spacings but changed which predicted structural positions were occupied. At the sequence wrap boundary, ordinary linear separations could change; the ≥20-residue eligibility criterion was recalculated for each shift. All nonzero shifts were used for lengths ≤1,200 aa; for longer models, 1,200 distinct shifts were drawn without replacement (per-gene seed 20261014 + sorted-gene index). A gene was testable only with ≥5 high-confidence missense sites, ≥10 eligible sequence-distant pairs and ≥100 valid shifts meeting those conditions. The empirical upper-tail p-value (1 + shifts with statistic ≥ observed)/(1 + valid shifts) was BH adjusted across all 74 testable genes. The remaining 225 received explicit no-model, too-few-sites, insufficient-confidence/pairs or insufficient-valid-shifts status. The cyclic shift largely retained one-dimensional clustering but did not preserve linear distances across its wrap boundary, domain-specific annotation, pLDDT distribution or interdomain alignment uncertainty; PAE was not evaluated. Structural proximity in the wild-type model was not interpreted as a mutant protein conformation or an LLPS phenotype.

Each gene was classified on two descriptive axes (relative C/H label-burden tier and detected/no detected/untestable sequence hotspot) with an independent 3D testability and adjusted-proximity flag. The burden thresholds were assigned within the mutation-ascertained source cohort. No additional 3D cluster class was assigned solely on unadjusted p-values. The per-gene classifications, count table and plot were retained alongside the variant-level coordinate table.

**Selection of partner-bound complexes and interface geometry**

For DLG4, CTNNB1, and HSPA8, partners were selected from the supplied union pair table only when a deposited partner-bound experimental structure could also be identified. The chosen structures were PDB **5JXB** (human DLG4 PDZ3 and a mouse SYNGAP C-terminal motif, 2.9 Å X-ray), **1JDH** (human CTNNB1 and TCF7L2 fragments, 1.9 Å X-ray), and **3LDQ** (human HSPA8 ATPase and BAG1 fragments, 1.9 Å X-ray). 5JXB was an engineered single-chain fusion; its DLG4 and mouse SYNGAP segments were split into two *display* chains while preserving deposited Cartesian coordinates. It was not represented as an independently assembled human–human dimer. The separate human fragments in 1JDH and 3LDQ were used in their deposited bound conformations. The small molecule in 3LDQ was omitted from protein–protein contact calculations. The separately deposited DLG4 Asp332Gly PDZ3 crystal structure (6QJK) was examined only as an **isolated mutant domain** and was not substituted for a mutant SYNGAP-bound complex.

Original PDB author residue numbers were transferred to UniProt positions using PDBe SIFTS mappings for the relevant protein chains. Mapped variant residues were accepted only if the residue identity in the experimental construct matched the source reference residue. For each accepted target residue, the shortest Euclidean distance in ångströms between any of its nonhydrogen atoms **and** any nonhydrogen atom of the bound partner protein was calculated on the original deposited three-dimensional coordinates. A site was classed as directly contacting at ≤5 Å, nearby at >5 to ≤10 Å, and not contacting at >10 Å; residues outside the resolved construct and complex labels without a unique site were recorded separately. Each unique target–partner residue contact within 5 Å was also tabulated. These were descriptive geometry cutoffs, not binding-energy, affinity, or functional-effect estimates. Cα traces were projected orthographically for figure display; no distance was inferred from the two-dimensional projection. Source labels at a shared residue were counted as one structural site when discussing the interface, and all substitutions at that site were retained in the underlying variant table.

**Statistical scope and reproducibility**

For multiple-hypothesis families described above, the specified BH correction was applied within the stated family, not to every descriptive count in the project. Monte Carlo p-value floors depended on the analysis: 1/100001 for the 100,000-draw source-compartment/DrLLPS tests; 1/20001 for the 20,000-draw 2025 LLPS role and selected-example tests; and 1/5001 for the cohort-wide sequence scan. The structural circular-shift test had a gene-specific number of valid shifts. Cross-cell reuse of the same proteins and edges, overlaps among libraries and constituent sets, and overlapping annotation databases precluded interpreting rows, cell types, or source records as independent biological replicates. The pair-incidence reconstruction, plateau annotations, variant mappings, residue-distance calculations, source construct notes, and all included/excluded cases were retained in companion CSV files. The files available for this study support the condensate, mutation, cell-conditioned PPI, and structural-interface sections above. The mouse–human connectome analyses, AHBA node analyses, TFBS/background analyses, regional synaptic-withdrawal controls, and details of other earlier HAP analyses are available in the prior publications (Glinsky, 2026a; b; c), including their source files, cohort definitions, nulls, parcellation mapping, and code, if applicable.

**Acknowledgements**

This is the single author contribution. All analyses were conceived, prompted and initiated by the author. Implementation and execution of the analyses were carried out with the assistance of Academia Co-scientist (Anthropic Claude, model Opus 5), operating on the primary data files under the author's specification; all reported values were verified by the author against those files. Academia Co-scientist (Anthropic Claude, model Opus 5; GPT-6 Sol) was used for computational analysis, figure generation and initial drafting of parts of the Introduction and Methods. All outputs were specified, verified and edited by the author, who takes full responsibility for the manuscript. The AI tool is not listed as an author and made no authorship-level contribution.

The authors used Academia Co-scientist (GPT-6 Sol; accessed October 2026) to assist with integrating public protein-annotation datasets, writing analysis code, and preparing preliminary draft figures and captions. The author used different AI assistants (Academia Co-scientist GPT-6 Sol and Anthropic Claude, model Opus 5) for independent cross-validation of the computational analyses outputs generated by the AI assistants. The reproducibility of the AI assistants' outputs starting the task of complex multi-step computational analyses of 154 human cell types from the independent retrievals of the public database records exceeds 99.5%. The authors retain responsibility for verifying the data, analyses, references, and final interpretation.

**Conflict of Interest**

The author declares that he has no conflicts of interest with the contents of this article.

**Figure legends.**

**Figure 1.** Whether the cascade reaches a protein set and what that set contains are set by different features: reach analysis by network structure and condensate content by branch. Eleven protein subsets of the 1,122-protein cascade network are compared on condensate-candidate content and on engagement by the cascade, which was run to its plateau separately in each of 154 HPA single-cell consensus cell types (availability nTPM ≥ 1; hypergeometric term test, Benjamini-Hochberg (BH) $q < 0.05$, iterated to convergence) over the Enrichr Transcription_Factor_PPIs (tftf) and PPI_Hub_Proteins (hub) libraries.
(**A**) Condensate content of each subset as z against 20,000 permutations of the condensate label within internal-degree deciles of the network; bars are red at $z \geq +2$ and blue at $z \leq -2$, and the reported value is the raw condensate rate. The 716 initiators (39.8%, z = +3.94) and the TF–TF family T245 (47.8%, z = +3.22) carry more condensate annotation than their connectivity predicts; the hub family H350 carries far less (26.9%, z = −5.77). The 11 subsets of the PPI cascade constituents ordered by z of condensate content against a degree-matched draw over the 1,122-protein network. The reported percentage is the raw condensate rate.
(**B**) Median internal degree against condensate rate, marker area proportional to √n, dashed line at the condensate rate of the network outside each subset. T245 and H350 are joined by a line: matched on connectivity (488 against 506) and 21 percentage points apart in condensate content.
(**C**) Fraction of each subset reached at the cascade plateau, median over the 154 cell types, against condensate z. Squares mark subsets that are under 50% library terms — the 716 initiators, P545 and C171 are 25.4%, 27.9% and 17.5% terms, so most of their members are seeds rather than destinations and cannot be reached at any expression level (shaded band, the region where reach is structurally capped), despite 56% of them being available. Circles mark subsets that are 100% library terms. Among subsets that are library terms, reach tracks degree (Spearman ρ = +0.764, p = 0.0062, n = 11). T245 PPI cascade family is the subset that is both fully reachable (66.5%) and condensate-enriched. The annotation points at T245, which is the one subset that is fully a term, reached in two thirds of cell types, and condensate-forming protein-enriched.
(**D**) Condensate rate of the seed layer, of the network, and of the terms reached at the plateau in each branch, as medians over the 154 cell types. The plateau is condensate-richer than the seed layer it grows from, and the TF-TF branch (52.1%) is richer than the hub branch (38.8%) in 154 of 154 cell types. Condensate membership is a candidate annotation, not a measurement of phase separation.

**Figure 2.** Unsupervised clustering of cell-type PPI cascade constituents-associated protein dimers' repertoires including in each protein dimer at least one monomer protein selected from candidates cascade-forming proteins. Each of 154 human cell types is represented by a binary profile of 51,439 non-self, unordered predicted protein dimer types under the supplied numerical ≥1 availability rule. The dendrogram uses average linkage on Jaccard distance (1 minus intersection/union); optimal leaf ordering changes display order. The adjacent heatmap shows cell-to-cell Jaccard similarity in the same leaf order (scale 0–1). Colored bands identify the reported in the main text five cell-type sets (46 highlighted cell types total). Four of these five sets are pure local subtrees in this dendrogram tree. The six spermatogenesis cell types appear as adjacent leaves but do not constitute a single pure subtree. The five colors are the selected examples of cell type annotations, not the five groups obtained by a global k = 5 cut, which yields one 150-cell group and four singleton outliers. Pair eligibility was inferred from expression availability and PPI-library term–member relationships. Protein

dimer formation remains to be experimentally verified. Table 3 reports top-scoring examples of diverse representations of the PPI cascade constituents-associated protein heterodimers with at least one monomer protein selected from candidate condensate-forming proteins of five groups of human cells defined by the unsupervised hierarchical clustering of 154 human cell types.

**Figure 3.** Phase-separated condensate-associated annotation of 154 human cell type-conditioned PPI repertoires of Hybrid Assembly Pathway (HAP) constituents.
(**A**) Observed-to-null ratios for heterodimers whose two endpoints share a phase-separated compartment annotation of Banani et al. (2022). The primary statistic weights each edge by the number of HPA cell types in which it was reconstructed. Whole 35-term annotation vectors were shuffled together among proteins matched by heterodimer-graph degree and HPA detection breadth (100,000 permutations), holding edges and their cell frequencies fixed; numbers to the right are Benjamini–Hochberg q-values across all 35 terms and distinct annotated edge counts. Teal and ochre indicate $q < 0.05$ for direct GO labels and functional GO proxies, respectively; gray marks nonsignificant controls. Marker area scales with distinct edge count. The same ten terms pass $q < 0.05$ when every distinct edge is counted once.
(**B**) Range and median of descriptive, cell-specific observed/degree-bin-expected pairing ratios for eight terms; the right column counts cells with a ratio >1. This expectation differs from the joint null in A; no cell-specific p-values are inferred from B.
(**C**) Fractions of reached TF–TF and hub plateau proteins bearing four annotations under the nTPM ≥1 rule; each faint dot is one of 154 cell types, with median diamonds. Panel (**C)** describes reached-protein content that are known to assemble dimers. † indicates a compartment name represented in the source by related functional GO terms rather than a matching localization term. The analysis includes 51,439 reconstructed heterodimer types and 5,919,637 cell–heterodimer occurrences; 200 discordant pair types were excluded. Shared endpoint annotation and PPI support reflect the likelihood, but do not establish, dimer formation inside a condensate or cause the phase separation. Primary data sources: Banani et al. (2022); Human Protein Atlas v25.1; Glinsky, 2026a; b. Edge evidence was described as experimentally supported based on primary publications but original provenance was not subjected to independent audit.

**Figure 4.** Pathogenic feature-overlap annotations in the cell-conditioned PPI cascade.
(**A**) Putative phase-separated condensate-forming proteins (Banani et al. (2022) were mapped to current human gene names (n = 3,932), and compared with those also represented in the 1,122-symbol cascade network (n = 409). Teal denotes genes with ≥1 pathogenic mutation mapped to a condensate-promoting feature; ochre denotes another catalogued pathogenic mutation but not condensate-promoting feature-overlap annotation; gray denotes neither. Counts are mutation-harboring genes, not independent mutation events.
(**B**) Two-sided Fisher odds ratios and 95% confidence intervals for annotation among mutation-ascertained candidate genes in the indicated families versus other mutated candidate genes. Displayed q-values apply Benjamini–Hochberg correction to the five raw contrasts. The separate network estimate adjusts by logistic regression for protein length, total catalogued pathogenic-mutation count, and MID annotation; its displayed p is unadjusted.
(**C**) Mutation-positive genes encoding condensate-forming proteins as a fraction of reached candidate proteins at the TF–TF and TF-HUB plateaus for 154 HPA human cell types under the two project availability rules. Faint points denote cells, thick denotes segments interquartile ranges, and diamonds denote medians.
(**D**) Among ten source GO compartments previously significant in a separate global PPI-pair co-annotation test (Figure 2), the percentage of GO-annotated cell–heterodimer occurrences with at least one condensate-forming protein mutation-positive endpoint. Right-hand counts are distinct interacting pair types with such an endpoint over all pair types annotated to the compartment. † A functionally related GO proxy rather than a direct compartment-localization term.
Panels C–D are descriptive and do not imply that HPA cells carry these variants or that the mutation variants affect their dimers. Mutations' annotation is a feature-overlap screen, not an allele-specific demonstration of condensate disruption. Independent self-assembly precedents are TP53 R337H, which destabilizes the p53 tetramer [DiGiammarino et al., Nat Struct Biol 9:12–16 (2002), doi:10.1038/nsb730], and STAT3 Y640F, which stabilizes activated homodimers [Mori et al., Sci Rep 7:15919 (2017), doi:10.1038/s41598-017-15912-6]. Sources: Banani et al., Developmental Cell 57:1776–1788.e8 (2022); HPA version 25.1; Glinsky, 2026a, b.

**Figure 5.** Curated liquid-liquid phase-separated (LLPS) condensate evidence refines the PPI cascade cascade pairing claims.
(**A**) Among 1,122 supplied cascade symbols (1,011 matching current HPA/UniProt primary human gene names), the role-labelled subsets contain 13 PhaSePro drivers, 18 DrLLPS scaffolds, 37 regulators and 166 potential clients. Colored portions count proteins with Banani et al. (2022) pathogenic mutations mapped to condensate-promoting features; each bar uses its own role-specific denominator.
(**B**) Same-role partner pairs in 51,439 reconstructed heterodimer types (5,919,637 inferred cell–pair occurrences in 154 human cell types). Observed counts are divided by the mean of 100,000 upper-tail joint-annotation permutations stratified by retained PPI-graph degree and across-cell HPA expression-detection breadth. Filled circles weight each inferred cell–pair occurrence; open circles count each distinct pair type once. Shown q-values are Benjamini–Hochberg adjusted over four role tests for the occurrence-weighted primary analysis (PhaSePro driver 0.000587; DrLLPS features: scaffold 0.000440; regulator 0.000240; client 0.632). The first three roles also pass the distinct-edge sensitivity test; client annotation does not.
(**C**) Within candidate genes with at least one catalogued pathogenic mutation, unadjusted two-sided Fisher odds ratios for feature-overlap mutation annotation are shown with 95% confidence intervals, comparing each evidence role to other mutation-ascertained candidates; q-values adjust four overlapping role tests. Diamonds show a separate binomial sensitivity model jointly including all four role indicators, protein length, total catalogued mutation count, and MID annotation.
(**D**) Six tracked PPI pairs matching PhaSePro-described partner-dependent LLPS systems; the x-axis gives inferred pair eligibility among 154 reference cell types, not observations of cellular dimers or condensate formation. DLG4–SYNGAP1 is a testable partner-specific case: the listed DLG4 Asp332Gly variant lies within a mapped PDZ interval, but its binding or condensation effect has not been tested. GRB2 and NCK1 pairs occur within multicomponent systems, not isolated binary-droplet demonstrations. The APP–PRNP graph edge is a proxy of and should not be equated to the amyloid-β-oligomer/PrPc species assayed in PhaSePro. The DrLLPS potential-client tier is not synonymous with validated LLPS drivers. PhaSePro was among the sources of Banani et al. (2022) candidate union, and shared ascertainment bias precludes interpreting database overlap as independent experimental replication.
Human PhaSePro entries were mapped by UniProt accession to experimentally curated LLPS-driver status; DrLLPS human entries supplied separate definitions for scaffold, regulator, and potential-client roles. Isoform suffixes were removed for current human gene mapping. Primary sources: Mészáros et al. (2020); Ning et al. (2020); Banani et al. (2022); HPA v25.1; Glinsky, 2026a, b. Primary partner studies linked in the Supplementary table phasepro_partner_dependent_edges.csv.

**Figure 6.** Synaptic condensate provenance identifies a prioritized PSD-95 partner route within the PPI cascade TF-HUB protein branch (H350 family of PPI cascade hub proteins).
(**A**) Successively more restrictive subsets of the 245 mapped PPI cascade TF-HUB constituents (hub proteins), displayed on a logarithmic count axis: any SynGO association (94 hub proteins); an explicit post-synaptic density (PSD) cellular-component (CC) annotation in GO:0014069, GO:0099092, GO:0099031, or GO:0099061 (23 hub proteins); the same PSD CC criterion plus a DrLLPS Post-synaptic density tag (12 hubs); and a PhaSePro experimentally curated PSD driver (one: DLG4/PSD-95). The separate count 74/245 with any DrLLPS PSD tag is not a step in the nested series; 63/74 are potential clients, which cannot be re-labelled as LLPS drivers.
(**B**) The 12 hub proteins meeting both explicit PSD CC and DrLLPS PSD criteria, retaining their distinct source roles: DLG4 and FYN (scaffolds), GSK3B (regulator), and nine potential clients. PhaSePro characterizes DLG4 phase separation as partner-dependent, not autonomous; its absence for another protein does not show that the protein cannot undergo LLPS. Banani et al. (2022) pathogenic mutations were mapped to condensate-promoting features, not a measured perturbation of the PSD. The last column counts cell types with a listed DLG4 heterodimer in the supplied PPI graph; a dash in the last column means no listed heterodimer or record available, not negative evidence.
(**C**) Candidate routes from DLG4, colored by external experimental context, with x-axis giving inferred pair eligibility among 154 reference-expression cell types. DLG4–SYNGAP1 (66/154 cell types) maps to an experimentally studied PSD-95/SynGAP co-assembly and PhaSePro partner dependency (Zeng et al. 2016). DLG4–DLGAP1/GKAP (142/154 cell types) participates in a multicomponent PSD reconstitution with Shank3 and Homer3 (Zeng et al. 2018), not a claim that the isolated DLG4–DLGAP1 heterodimer alone phase separates. SYNGAP1 and DLGAP1 are outside the supplied 1,122-symbol family list, but their pair types occur in the broader PPI graph as available protein monomers. FYN, GRIN1 and CAMK2A are H350 proteins with tracked DLG4 pairs.

(**D**) Source-reported DLG4 PDZ and SH3 intervals on its 724-aa sequence; the Asp332Gly mutation feature-overlap allele lies within the annotated third PDZ interval (314–394). This supports an experimentally-testable variant–partner hypothesis, it is not an explicit report of an observed effect on SynGAP binding, PSD partitioning or condensate formation.

The supplied H350 list comprises 350 symbols, 245 of which map to current HPA/UniProt human primary gene symbols. Human SynGO annotations were taken from the Harmonizome 3.0 processed gene–term table (last updated March 2024), and DrLLPS and PhaSePro records were mapped by UniProt accession.

Primary sources: SynGO Gene Annotations dataset: SynGO-derived Harmonizome 2024 gene–term table; Liquid-liquid phase separation (LLPS) databases: PhaSePro; DrLLPS; Banani et al. (2022); Zeng et al., *Cell* 166 (2016); Zeng et al., *Cell* 174 (2018). Glinsky, 2026a, b.

**Figure 7.** Candidate condensate-feature mutation variants relative to three partner-bound interfaces.

Upper 3 panels show orthographic projections of Cα traces from experimentally determined partner-bound fragments: human DLG4 PDZ3 with a mouse SYNGAP C-terminal motif (PDB 5JXB; 2.9 Å); human CTNNB1 with TCF7L2 (1JDH; 1.9 Å); and human HSPA8 ATPase domain with BAG1 (3LDQ; 1.9 Å). The 5JXB construct is an engineered fusion; its DLG4 and mouse SYNGAP segments are displayed as separate chains for clarity and must not be interpreted as a freely assembled human heterodimer. Small cyan and gold points mark target and partner residues with an interchain protein-heavy-atom contact ≤5 Å. Large colored points mark UniProt-numbered target sites carrying mutations reported in Banani et al. (2022), a computational catalog of pathogenic variants overlapping putative condensate-promoting features. Labels at the same residue count as one site.

Lower panel gives the shortest three-dimensional heavy-atom distance from each highlighted target residue to the partner, measured in the deposited reference complex (not from the 2D projection). CTNNB1 Lys335 (2.89 Å; two catalogued substitutions); Lys292 (3.47 Å; three); and Asn387 (3.75 Å; four) make direct contact with TCF7L2. DLG4 Asp332Gly is 8.09 Å from the SYNGAP motif and therefore classified as near, but not at, the defined direct interface. HSPA8 Ala133Ser lies 22.96 Å from BAG1; no HSPA8 mutation site contacts BAG1 in this fragment. Other DLG4 mutation sites (Gly198, Val476, Tyr50 frameshift start) and HSPA8 mutation sites (Thr429, Gln473) fall outside their respective resolved constructs; many CTNNB1 N-terminal sites are likewise not visualized. The 5 Å and 10 Å thresholds are descriptive geometric bins, not affinity or functional cutoffs.

The reference PDB wild-type protein geometry is not a mutant structural model. Interface location offers a testable mechanism for the three CTNNB1 sites, not evidence that those substitutions alter binding or condensate formation; the two other tested partner interfaces do not show direct mutation-site overlap. Bound coordinates are from RCSB PDB 5JXB, 1JDH, and 3LDQ. PDB numbering (Banani et al., 2022) were transferred to UniProt reference positions with PDBe SIFTS. Cα traces are projected for display; the distances, on the original un-projected coordinates, use all resolved protein heavy atoms. The computed table dimer_census_union.csv includes the selected pairs in 66/154, 153/154, and 154/154 cell types of the HPA panel, respectively. Conclusions are restricted to the observed fragment, partner and a structural state. An isolated DLG4 D332G mutant PDZ3 crystal structure (6QJK) is available separately, but contains no bound SYNGAP.

**Table 1.** Enrichment analysis of the PPI cascade constituents* among the 3,862 proteins defined as the candidate protein components of phase-separated condensates.

| Background | Background rate | Network rate** | Ratio, O/E | Number, expected | P value |
|---|---|---|---|---|---|
| **Human proteome (20,190)** | 19.1% (3862/20190) | 40.5% (409/1,011) | 2.12 | 193.4 | 1.4e-58 |
| **HPA***-measured genes (9,701)** | 27.1% | 40.5% | 1.49 | 273.6 | 1.0e-22 |
| **Proteome ∩ HPA-measured (9,635)** | 27.2% | 40.5% | 1.49 | 275.4 | 4.1e-22 |

Legend: *, 1,011 of the 1,122 HAP constituents map into the reviewed proteome; **, 409 of 1,011 HAP constituents overlap 3,862 condensate proteins; O/E, observed/expected ratio; HPA***, Human Protein Atlas; ∩ = intersection. 3,868 of the 3,941 entry names (Banani et al., 2022) carry a primary gene symbol, giving 3,862 distinct symbols of proteins defined as the candidate protein components of phase-separated condensates.

**Table 2.** Comparisons of analytical metrics of the PPI cascade constituents.

| PPI cascade subset | n | condensate % | z | median degree | % library terms | % reached | tftf / hub reach | % available |
|---|---|---|---|---|---|---|---|---|
| **716 initiators** | 716 | 39.8 | +3.94 | 178 | 25.4 | 15.1 | 14.7 / 2.8 | 56.1 |
| **T245 TF–TF family** | 245 | 47.8 | +3.22 | 488 | 100 | 66.5 | 66.1 / 11.8 | 80.0 |
| **P545 principal initiators** | 545 | 40.7 | +3.01 | 249 | 27.9 | 16.5 | 16.1 / 2.9 | 55.8 |
| **C171 initiators complementary** | 171 | 36.8 | +0.53 | 30 | 17.5 | 11.1 | 9.9 / 2.3 | 56.6 |
| **128 proteinopathy-bound targets** | 128 | 36.7 | −1.36 | 586 | 100 | 85.9 | 25.8 / 68.8 | 92.1 |
| **81 SynGO hub proteins** | 81 | 29.6 | −2.53 | 580 | 100 | 93.8 | 3.7 / 93.8 | 95.1 |
| **H350 hub family** | 350 | 26.9 | −5.77 | 506 | 100 | 62.0 | 8.0 / 62.0 | 92.7 |

**Table 4.** At the plateau the branches separate in both labels, and in the same direction in every cell type (medians over 154 cell types).

| PPI cascade stage | CORUM % | condensate % | complexes per member |
|---|---|---|---|
| **seed layer, 716 initiators** | 14.5 | 39.8 | 0 |
| **network 1122** | 25.1 | 36.5 | — |
| **tftf branch at plateau** | 57.7 | **52.1** | 1 |
| **hub branch at plateau** | **65.8** | 38.8 | 2 |

**Table 5.** Statistical metrics of overrepresented 10 of 35 phase-separated condensate compartment annotations.

| Phase-separation Compartment | Annotated endpoints | Observed cell–pair occurrences | Null mean | Observed/null | Permutation p | BH q (35) |
|---|---|---|---|---|---|---|
| **DNA repair focus†** | 152 | 28,018 | 6,951 | 4.03 | $1.0×10^{-5}$ | $8.75×10^{-5}$ |
| **PML body** | 56 | 14,243 | 2,649 | 5.38 | $1.0×10^{-5}$ | $8.75×10^{-5}$ |
| **chromatin** | 264 | 170,020 | 53,315 | 3.19 | $1.0×10^{-5}$ | $8.75×10^{-5}$ |
| **transcriptional condensate†** | 807 | 562,067 | 270,864 | 2.08 | $1.0×10^{-5}$ | $8.75×10^{-5}$ |
| **Cajal body** | 24 | 1,254 | 77 | 16.27 | $4.00×10^{-5}$ | $2.80×10^{-4}$ |
| **centrosome** | 344 | 46,878 | 25,438 | 1.84 | 0.00127 | 0.00741 |
| **nuclear body** | 504 | 110,156 | 68,819 | 1.60 | 0.00166 | 0.00830 |
| **DNA replication condensate†** | 47 | 1,643 | 294 | 5.59 | 0.00217 | 0.00914 |
| **nucleolus** | 557 | 103,235 | 65,101 | 1.59 | 0.00245 | 0.00914 |
| **focal adhesion** | 316 | 71,423 | 43,509 | 1.64 | 0.00261 | 0.00914 |

**Table 6.** Identification of the candidate condensate-disrupting pathogenic mutations among the PPI cascade constituents*.

| Set | Feature-overlap + / mutation-ascertained | Comparator + / mutation-ascertained | OR | BH q |
|---|---|---|---|---|
| **P545** | 154/168 | 1,587/2,199 | 4.24 | $4.45\times10^{-9}$ |
| **C171** | 37/40 | 1,704/2,327 | 4.51 | 0.00452 |
| **T245** | 90/96 | 1,651/2,271 | 5.63 | $6.40\times10^{-7}$ |
| **H350** | 57/70 | 1,684/2,297 | 1.60 | 0.168 |
| **All mapped network candidates** | 269/299 | 1,472/2,068 | 3.63 | $6.25\times10^{-13}$ |

Legend: *, Values are gene counts, not mutation-event counts. BH correction covers these four family tests plus the whole-network contrast (five tests).

**Table 7.** Examples of the PPI cascade proteins TP53 and STAT3 condensate-disrupting mutations affecting self-association

| Source variant | Matched feature and tracked self-pair | Independent evidence | What it does not demonstrate |
|---|---|---|---|
| **TP53 Arg337His** | p53 tetramerisation domain (319–357); TP53–TP53 is in the union in 152/154 cell types | DiGiammarino et al., *Nature Structural Biology* 9:12–16 (2002), doi:10.1038/nsb730, demonstrated **pH-dependent destabilization of the mutant p53 tetramer**. | A change to any particular TP53 **heterodimer**, such as TP53–MDM2 (a separate tracked pair). |
| **STAT3 Tyr640Phe** | SH2 domain; STAT3–STAT3 is in the union in 154/154 cell types | Mori et al., *Scientific Reports* 7:15919 (2017), doi:10.1038/s41598-017-15912-6, reported **stabilization of activated STAT3 homodimers**. Pilati et al., *Journal of Experimental Medicine* 208:1359–1366 (2011), doi:10.1084/jem.20110283, also investigated activating STAT3 mutations in hepatocellular adenoma. | A direction or magnitude for STAT3–STAT1 or any other specific heterodimer, or condensate occupancy. |

**Table 8.** The global pair test on the 51,439 reconstructed heterodimer types spanning 5,919,637 cell–pair occurrences annotating endpoints by four distinct LLPS evidence roles.

| Endpoint label on both partners | Human role-labeled union endpoints | Observed cell–pair occurrences | Observed / null mean | BH q | Distinct pair types | Distinct-edge BH q |
|---|---|---|---|---|---|---|
| **PhaSePro curated driver** | 54(59) | 8,328 | 2.89 | 0.000587 | 63 | 0.000693 |
| **DrLLPS scaffold** | 80(86) | 17,197 | 2.51 | 0.000440 | 123 | 0.000693 |
| **DrLLPS regulator** | 290(357) | 61,640 | 2.02 | 0.000240 | 463 | 0.000440 |
| **DrLLPS potential client** | 2,453(3,190) | 1,004,795 | 0.974 | 0.632 | 7,823 | 0.657 |

Legend: The driver/scaffold results survive distinct-edge weighting, but shared citation, assay and protein-network topology beyond degree bins remain possible explanations.

**Table 9.** Cross-database evidence ladder of the PPI cascade family proteins' engagements in post-synaptic density (PSD) phase-separated condensates.

| Evidence level among H350 family proteins | Count | Meaning |
|---|---|---|
| **SynGO synaptic genes, any term** | 94 | Expert-curated synaptic association; not an LLPS annotation. |
| **SynGO explicit PSD cell content term** | 23 | PSD location or membrane subcomponent. |
| **DrLLPS records with PSD tag** | 74 | 63 clients, 7 regulators, 4 scaffolds; **not** 74 experimentally established PSD drivers. |
| **Both any SynGO term and DrLLPS PSD tag** | 45 | Synaptic association plus a condensate-database PSD label; sources may share evidence. |
| **Both explicit SynGO PSD cell content and DrLLPS PSD tag** | **12** | Highest-precision location × database shortlist; individual role strength differs. |
| **PhaSePro experimentally curated PSD driver** | **1: DLG4** | PhaSePro documents PSD-95/SynGAP partner-dependent phase separation. |

**Table 10.** Summary of the statistical significance evaluations of the interrogations of the reconstructed protein heterodimer graphs across the 154 human cell types.

| Both endpoints annotated | Labeled graph endpoints | Cell–pair occurrences, observed / null | Weighted BH q | Distinct edge types, observed / null | Edge BH q |
|---|---|---|---|---|---|
| **Any positive** | 320 (382) | 125,447 / 44,532 (2.82×) | 0.000125 | 998 / 358.1 (2.79×) | 0.000125 |
| **DE** | 129 (146) | 21,543 / 8,211 (2.62×) | 0.000417 | 171 / 65.8 (2.60×) | 0.000167 |
| **CE** | 157 (199) | 16,228 / 8,015 (2.02×) | 0.0114 | 125 / 66.5 (1.88×) | 0.0168 |
| **C_D** | 34 (37) | 5,895 / 884 (6.67×) | 0.000125 | 47 / 6.56 (7.16×) | 0.000125 |
| **D+** | 48 (50) | 5,667 / 1,393 (4.07×) | 0.000562 | 43 / 10.9 (3.94×) | 0.000562 |

Legend: Numbers in parenthesis are the numbers of all human proteins assigned to the corresponding graph endpoints.

**Table 11.** Overlaps of all five classifications of condensate-positive proteins with PPI cascade constituents are significant under the genome-wide-human reference after correction for five tests.

| Annotation | Graph endpoints / identified humans | Expected endpoints | Observed / expected | Odds ratio (exact 95% CI) | Two-sided Fisher p | Holm-adjusted p |
|---|---|---|---|---|---|---|
| **Any positive** | 320 (382) | 166.9 | 1.92× | 6.86 (5.21–9.17) | $7.81 \times 10^{-60}$ | $\mathbf{3.91 \times 10^{-59}}$ |
| **DE** | 129 (146) | 63.8 | 2.02× | 9.91 (5.95–17.55) | $1.32 \times 10^{-29}$ | $\mathbf{5.29 \times 10^{-29}}$ |
| **CE** | 157 (199) | 87.0 | 1.81× | 4.89 (3.45–7.05) | $3.20 \times 10^{-24}$ | $\mathbf{9.59 \times 10^{-24}}$ |
| **C_D** | 34 (37) | 16.2 | 2.10× | 14.66 (4.61–74.60) | $1.43 \times 10^{-9}$ | $\mathbf{1.43 \times 10^{-9}}$ |
| **D+** | 48 (50) | 21.8 | 2.20× | 31.09 (8.15–264.12) | $2.06 \times 10^{-15}$ | $\mathbf{4.12 \times 10^{-15}}$ |

Legend: Numbers in parenthesis are the numbers of all human proteins assigned to the corresponding graph endpoints. D+ classification overlaps refer to the other condensate-positive annotations, so the rows are not independent categories.

**Table 12.** Summary statistic metrics of the 3D modelling using the AlphaFold DB run 2026-10-05.

| Measure | Result |
|---|---|
| **Genes in requested intersection** | 299 |
| **Source accession records** | 300 |
| **AlphaFold DB wild-type models retrieved** | 298 records / 297 genes |
| **Genes with ≥1 sequence-validated site** | 295 |
| **Unique mapped accession–residue sites** | 12,861 |
| **Pathogenic-label records parsed from S4C/S4H*** | 22,195 |
| **Successfully mapped records** | 17,947 |
| **S4H feature-overlap records mapped*** | 9,038 of 12,237 |
| **S4H missense records mapped*** | 4,493 |
| **Mapped H labels at local pLDDT < 70** | 4,266 of 9,038 |

Legend: *, human pathogenic mutation label records were parsed from the Supplementary Table S4C and S4H reported in the Banani et al. (2022). Each label is checked against the 2020_06 source sequence, current AlphaFold DB reference sequence, model fragment boundaries, and the PDB Cα residue identity. Only labels passing every check get coordinates and pLDDT. For missense and single-residue deletion/duplication/stop labels, the coordinate is the wild-type residue. For frameshifts, it is the start of the frameshift, not the altered downstream chain. Range indels, unknown starts, extensions and other non-single-site notations are not given an arbitrary single coordinate. Duplicate variants at one site remain separate records in the CSV; colored markers represent unique sites, not events. APC (P25054; 2,843 amino acids, 2,221 mutation labels) and HTT (P42858; 3,142 amino acids, two mutation labels) have no AlphaFold DB prediction at these accessions. They are not presented as modeled. RFX7 and SRRM2 have an available model but none of their source labels passes the site/fragment/reference checks; their wild-type 3D models are in the archive but there is no mapped colored variant site.

**Table 13.** Human disease-associated mutations relevant to the experimentally-validated binding interface of DLG4–SYNGAP1, CTNNB1–TCF7L2, and HSPA8–BAG1 dimers.

| Target–partner | Experimental structure | Structural ascertainment caveats | Candidate-pair cell types |
|---|---|---|---|
| **DLG4–SYNGAP1** | 5JXB, X-ray, 2.9 Å | Human DLG4 PDZ3 bound to a mouse SYNGAP C-terminal motif; crystallized as an engineered single- | 66/154 |
| | | chain fusion, split into two display chains. Not a free human–human dimer. | |
| **CTNNB1–TCF7L2** | 1JDH, X-ray, 1.9 Å | Separate bound human fragments: CTNNB1 armadillo region and a TCF7L2 peptide. | 153/154 |
| **HSPA8–BAG1** | 3LDQ, X-ray, 1.9 Å | Separate human HSPA8 ATPase and BAG1 BAG-domain fragments; structure also contains a small molecule, excluded from protein–protein distance measurements. | 154/154 |

**Table 3.** Diverse representations of the PPI cascade constituents-associated heterodimers with at least one monomer protein selected from candidate condensate-forming proteins of five groups of human cells defined by the unsupervised hierarchical clustering of 154 human cell types.

| Genes availability rule | Example group (G1-G5) | Diverse dimer rank | gene A | gene B | UniProt A | UniProt B | group cells with . | n group | other cells with . | n other | prevalence difference | BH q all five groups | also enriched other threshold |
|---|---|---|---|---|---|---|---|---|---|---|---|---|---|
| **ntpm>=1** | G1 spermatogenesis | 1 | BRCA1 | CETN1 | P38398 | Q12798 | 6 | 6 | 3 | 148 | 0.97973 | 3.04E-06 | 1 |
| **ntpm>=1** | G1 spermatogenesis | 2 | DMRTC2 | NR3C1 | Q8IXT2 | P04150 | 6 | 6 | 3 | 148 | 0.97973 | 3.04E-06 | 1 |
| **ntpm>=1** | G1 spermatogenesis | 3 | ESR2 | RBMXL2 | Q92731 | O75526 | 6 | 6 | 3 | 148 | 0.97973 | 3.04E-06 | 1 |
| **ntpm>=1** | G1 spermatogenesis | 4 | GATA1 | TRIM29 | P15976 | Q14134 | 6 | 6 | 3 | 148 | 0.97973 | 3.04E-06 | 1 |
| **ntpm>=1** | G1 spermatogenesis | 5 | ACTB | PFN3 | P60709 | P60673 | 6 | 6 | 4 | 148 | 0.972973 | 5.73E-06 | 1 |
| **ntpm>=1** | G1 spermatogenesis | 6 | DMRTB1 | EWSR1 | Q96MA1 | Q01844 | 6 | 6 | 4 | 148 | 0.972973 | 5.73E-06 | 1 |
| **ntpm>=1** | G1 spermatogenesis | 7 | CTCFL | SP1 | Q8NI51 | P08047 | 6 | 6 | 5 | 148 | 0.966216 | 1.07E-05 | 1 |
| **ntpm>=1** | G1 spermatogenesis | 8 | DNAJB8 | HDAC4 | Q8NHS0 | P56524 | 6 | 6 | 6 | 148 | 0.959459 | 1.71E-05 | 1 |
| **ntpm>=1** | G1 spermatogenesis | 9 | DDX53 | HDAC2 | Q86TM3 | Q92769 | 6 | 6 | 7 | 148 | 0.952703 | 2.87E-05 | 1 |
| **ntpm>=1** | G1 spermatogenesis | 10 | ESR1 | TUBA3C | P03372 | P0DPH7 | 6 | 6 | 7 | 148 | 0.952703 | 2.87E-05 | 1 |
| **ntpm>=1** | G1 spermatogenesis | 11 | GABARAP | RPL10L | O95166 | Q96L21 | 6 | 6 | 8 | 148 | 0.945946 | 4.20E-05 | 1 |
| **ntpm>=1** | G1 spermatogenesis | 12 | ACTG1 | CAPZA3 | P63261 | Q96KX2 | 6 | 6 | 11 | 148 | 0.925676 | 0.000135 | 1 |
| **ntpm>=1** | G2 brain cells | 1 | GP1BB | PRKACA | P13224 | P17612 | 10 | 10 | 3 | 144 | 0.979167 | 6.66E-10 | 1 |
| **ntpm>=1** | G2 brain cells | 2 | EEF1G | SCHIP1 | P26641 | P0DPB3 | 10 | 10 | 4 | 144 | 0.972222 | 2.05E-09 | 1 |
| **ntpm>=1** | G2 brain cells | 3 | ACTB | BBS1 | P60709 | Q8NFJ9 | 10 | 10 | 6 | 144 | 0.958333 | 1.46E-08 | 1 |
| **ntpm>=1** | G2 brain cells | 4 | ALDOC | GRIN2B | P09972 | Q13224 | 10 | 10 | 7 | 144 | 0.951389 | 3.33E-08 | 1 |
| **ntpm>=1** | G2 brain cells | 5 | ATF2 | DDX39B | P15336 | Q13838 | 10 | 10 | 11 | 144 | 0.923611 | 3.03E-07 | 1 |
| **ntpm>=1** | G2 brain cells | 6 | DLG4 | NLGN2 | P78352 | Q8NFZ4 | 10 | 10 | 11 | 144 | 0.923611 | 3.03E-07 | 1 |
| **ntpm>=1** | G2 brain cells | 7 | ESR1 | RBM14 | P03372 | Q96PK6 | 10 | 10 | 11 | 144 | 0.923611 | 3.03E-07 | 1 |
| **ntpm>=1** | G2 brain cells | 8 | GAP43 | GRIN1 | P17677 | Q05586 | 10 | 10 | 12 | 144 | 0.916667 | 4.44E-07 | 1 |
| **ntpm>=1** | G2 brain cells | 9 | IRF9 | STAT5B | Q00978 | P51692 | 10 | 10 | 12 | 144 | 0.916667 | 4.44E-07 | 1 |
| **ntpm>=1** | G2 brain cells | 10 | BCL2 | MOAP1 | P10415 | Q96BY2 | 10 | 10 | 13 | 144 | 0.909722 | 6.80E-07 | 1 |
| **ntpm>=1** | G2 brain cells | 11 | EIF4A1 | GABARAP | P60842 | O95166 | 10 | 10 | 14 | 144 | 0.902778 | 1.08E-06 | 1 |
| **ntpm>=1** | G2 brain cells | 12 | CRK | PCDHA6 | P46108 | Q9UN73 | 9 | 10 | 0 | 144 | 0.9 | 3.52E-10 | 1 |
| **ntpm>=1** | G3 pituitary retina cells | 1 | FYN | GABRG2 | P06241 | P18507 | 8 | 8 | 22 | 146 | 0.849315 | 0.000159 | 1 |
| **ntpm>=1** | G3 pituitary retina cells | 2 | ACTL6B | SMARCA4 | O94805 | P51532 | 8 | 8 | 23 | 146 | 0.842466 | 0.000194 | 1 |
| **ntpm>=1** | G3 pituitary retina cells | 3 | INA | SNCA | Q16352 | P37840 | 8 | 8 | 26 | 146 | 0.821918 | 0.000391 | 1 |
| **ntpm>=1** | G3 pituitary retina cells | 4 | CACNG2 | MAPK9 | Q9Y698 | P45984 | 8 | 8 | 27 | 146 | 0.815068 | 0.000479 | 1 |
| **ntpm>=1** | G3 pituitary retina cells | 5 | GABARAPL2 | TTBK1 | P60520 | Q5TCY1 | 8 | 8 | 31 | 146 | 0.787671 | 0.000995 | 1 |
| **ntpm>=1** | G3 pituitary retina cells | 6 | PIK3R1 | SCG2 | P27986 | P13521 | 8 | 8 | 33 | 146 | 0.773973 | 0.00141 | 1 |
| **ntpm>=1** | G3 pituitary retina cells | 7 | DPYSL5 | HSPA5 | Q9BPU6 | P11021 | 8 | 8 | 35 | 146 | 0.760274 | 0.001901 | 1 |
| **ntpm>=1** | G3 pituitary retina cells | 8 | CALY | PRKCA | Q9NYX4 | P17252 | 8 | 8 | 38 | 146 | 0.739726 | 0.003114 | 1 |
| **ntpm>=1** | G3 pituitary retina cells | 9 | CHD3 | CHD5 | Q12873 | Q8TDI0 | 8 | 8 | 40 | 146 | 0.726027 | 0.003911 | 1 |
| **ntpm>=1** | G3 pituitary retina cells | 10 | JPH3 | SMAD3 | Q8WXH2 | P84022 | 8 | 8 | 41 | 146 | 0.719178 | 0.004479 | 1 |
| **ntpm>=1** | G3 pituitary retina cells | 11 | SMAD2 | TCEAL6 | Q15796 | Q6IPX3 | 7 | 8 | 27 | 146 | 0.690068 | 0.006028 | 1 |
| **ntpm>=1** | G3 pituitary retina cells | 12 | ATCAY | UBA1 | Q86WG3 | P22314 | 8 | 8 | 46 | 146 | 0.684932 | 0.008425 | 1 |
| **ntpm>=1** | G4 immune vascular cells | 1 | CSNK2A2 | STATH | P19784 | P02808 | 14 | 15 | 13 | 139 | 0.839808 | 2.80E-08 | 1 |
| **ntpm>=1** | G4 immune vascular cells | 2 | CD38 | LCK | P28907 | P06239 | 15 | 15 | 28 | 139 | 0.798561 | 6.32E-07 | 1 |
| **ntpm>=1** | G4 immune vascular cells | 3 | BATF | IRF4 | Q16520 | Q15306 | 14 | 15 | 19 | 139 | 0.796643 | 4.44E-07 | 1 |
| **ntpm>=1** | G4 immune vascular cells | 4 | IL2RA | PRKCB | P01589 | P05771 | 14 | 15 | 20 | 139 | 0.789448 | 6.70E-07 | 1 |
| **ntpm>=1** | G4 immune vascular cells | 5 | IL7R | PLEK | P16871 | P08567 | 14 | 15 | 24 | 139 | 0.760671 | 2.90E-06 | 1 |
| **ntpm>=1** | G4 immune vascular cells | 6 | CD2 | FYN | P06729 | P06241 | 13 | 15 | 22 | 139 | 0.708393 | 1.52E-05 | 1 |
| **ntpm>=1** | G4 immune vascular cells | 7 | BTK | WAS | Q06187 | P42768 | 13 | 15 | 24 | 139 | 0.694005 | 2.86E-05 | 1 |
| **ntpm>=1** | G4 immune vascular cells | 8 | LYN | SLAMF1 | P07948 | Q13291 | 12 | 15 | 15 | 139 | 0.692086 | 9.83E-06 | 1 |
| **ntpm>=1** | G4 immune vascular cells | 9 | IL10RA | PTPN6 | Q13651 | P29350 | 15 | 15 | 43 | 139 | 0.690647 | 3.22E-05 | 1 |
| **ntpm>=1** | G4 immune vascular cells | 10 | GRB2 | HP | P62993 | P00738 | 14 | 15 | 35 | 139 | 0.681535 | 6.29E-05 | 1 |
| **ntpm>=1** | G4 immune vascular cells | 11 | ABI3 | PAK1 | Q9P2A4 | Q13153 | 13 | 15 | 26 | 139 | 0.679616 | 5.14E-05 | 1 |
| **ntpm>=1** | G4 immune vascular cells | 12 | RUNX3 | VDR | Q13761 | P11473 | 15 | 15 | 46 | 139 | 0.669065 | 6.16E-05 | 1 |
| **ntpm>=1** | G5 hematopoiesis | 1 | DYNLL1 | MPO | P63167 | P05164 | 7 | 7 | 12 | 147 | 0.918367 | 3.76E-05 | 1 |
| **ntpm>=1** | G5 hematopoiesis | 2 | DPPA4 | VIM | Q7L190 | P08670 | 7 | 7 | 16 | 147 | 0.891156 | 0.000129 | 1 |
| **ntpm>=1** | G5 hematopoiesis | 3 | MAPK8 | NFE2 | P45983 | Q16621 | 7 | 7 | 20 | 147 | 0.863946 | 0.000354 | 1 |
| **ntpm>=1** | G5 hematopoiesis | 4 | BTK | WAS | Q06187 | P42768 | 7 | 7 | 30 | 147 | 0.795918 | 0.002375 | 1 |
| **ntpm>=1** | G5 hematopoiesis | 5 | CD84 | PTPN11 | Q9UIB8 | Q06124 | 7 | 7 | 42 | 147 | 0.714286 | 0.011369 | 1 |
| **ntpm>=1** | G5 hematopoiesis | 6 | GNAI2 | RGS18 | P04899 | Q9NS28 | 6 | 7 | 22 | 147 | 0.707483 | 0.0075 | 1 |
| **ntpm>=1** | G5 hematopoiesis | 7 | ELF1 | MLC1 | P32519 | Q15049 | 7 | 7 | 47 | 147 | 0.680272 | 0.020316 | 1 |
| **ntpm>=1** | G5 hematopoiesis | 8 | PLEK | PRKCA | P08567 | P17252 | 7 | 7 | 49 | 147 | 0.666667 | 0.023977 | 1 |
| **ntpm>=1** | G5 hematopoiesis | 9 | ABL1 | SPTA1 | P00519 | P02549 | 5 | 7 | 10 | 147 | 0.646259 | 0.005087 | 1 |
| **ntpm>=1** | G5 hematopoiesis | 10 | NOTCH1 | PIK3CG | P46531 | P48736 | 6 | 7 | 32 | 147 | 0.639456 | 0.032618 | 1 |
| **ntpm>=1** | G5 hematopoiesis | 11 | LMO2 | LYL1 | P25791 | P12980 | 7 | 7 | 54 | 147 | 0.632653 | 0.039681 | 1 |
| **ntpm>=1** | G5 hematopoiesis | 12 | LST1 | PRKCB | O00453 | P05771 | 6 | 7 | 34 | 147 | 0.62585 | 0.041068 | 1 |
| **top 4K** | G1 spermatogenesis | 1 | DYNLL1 | TUBA3C | P63167 | P0DPH7 | 6 | 6 | 0 | 148 | 1 | 8.37E-08 | 1 |
| **top 4K** | G1 spermatogenesis | 2 | FKBP6 | HSP90AA1 | O75344 | P07900 | 6 | 6 | 1 | 148 | 0.993243 | 3.43E-07 | 1 |
| **top 4K** | G1 spermatogenesis | 3 | TUBA3E | YWHAE | Q6PEY2 | P62258 | 5 | 6 | 1 | 148 | 0.826577 | 1.45E-05 | 1 |
| **top 4K** | G1 spermatogenesis | 4 | TUBA3D | YWHAQ | P0DPH8 | P27348 | 5 | 6 | 3 | 148 | 0.813063 | 7.48E-05 | 1 |
| **top 4K** | G1 spermatogenesis | 5 | BRCA1 | TEX101 | P38398 | Q9BY14 | 4 | 6 | 0 | 148 | 0.666667 | 9.88E-05 | 1 |
| **top 4K** | G1 spermatogenesis | 6 | DMRTC2 | NR3C1 | Q8IXT2 | P04150 | 4 | 6 | 0 | 148 | 0.666667 | 9.88E-05 | 1 |
| **top 4K** | G1 spermatogenesis | 7 | GABARAPL2 | RBMXL2 | P60520 | O75526 | 4 | 6 | 0 | 148 | 0.666667 | 9.88E-05 | 1 |
| **top 4K** | G1 spermatogenesis | 8 | PDCL2 | YWHAB | Q8N4E4 | P31946 | 4 | 6 | 0 | 148 | 0.666667 | 9.88E-05 | 1 |
| **top 4K** | G1 spermatogenesis | 9 | PPARD | TNP1 | Q03181 | P09430 | 4 | 6 | 0 | 148 | 0.666667 | 9.88E-05 | 1 |
| **top 4K** | G1 spermatogenesis | 10 | DMRTB1 | EWSR1 | Q96MA1 | Q01844 | 4 | 6 | 1 | 148 | 0.65991 | 0.000355 | 1 |
| **top 4K** | G1 spermatogenesis | 11 | BTG4 | UBC | Q9NY30 | P0CG48 | 4 | 6 | 2 | 148 | 0.653153 | 0.000854 | 1 |
| **top 4K** | G1 spermatogenesis | 12 | CHD5 | HDAC2 | Q8TDI0 | Q92769 | 4 | 6 | 12 | 148 | 0.585586 | 0.028497 | 1 |
| **top 4K** | G2 brain cells | 1 | ATF2 | DDX39B | P15336 | Q13838 | 10 | 10 | 0 | 144 | 1 | 4.13E-12 | 1 |
| **top 4K** | G2 brain cells | 2 | CSNK1E | EIF4A1 | P49674 | P60842 | 10 | 10 | 0 | 144 | 1 | 4.13E-12 | 1 |
| **top 4K** | G2 brain cells | 3 | EP300 | RBM14 | Q09472 | Q96PK6 | 10 | 10 | 0 | 144 | 1 | 4.13E-12 | 1 |
| **top 4K** | G2 brain cells | 4 | MDM2 | TSNAX | Q00987 | Q99598 | 10 | 10 | 0 | 144 | 1 | 4.13E-12 | 1 |
| **top 4K** | G2 brain cells | 5 | CD99L2 | SDHA | Q8TCZ2 | P31040 | 10 | 10 | 2 | 144 | 0.986111 | 2.35E-10 | 1 |
| **top 4K** | G2 brain cells | 6 | MAP1LC3B | SYNCRIP | Q9GZQ8 | O60506 | 10 | 10 | 5 | 144 | 0.965278 | 4.99E-09 | 1 |
| **top 4K** | G2 brain cells | 7 | HIF1A | ZNF197 | Q16665 | O14709 | 10 | 10 | 7 | 144 | 0.951389 | 2.66E-08 | 1 |
| **top 4K** | G2 brain cells | 8 | ATN1 | ITCH | P54259 | Q96J02 | 10 | 10 | 8 | 144 | 0.944444 | 4.34E-08 | 1 |
| **top 4K** | G2 brain cells | 9 | CALM1 | SIRT2 | P0DP23 | Q8IXJ6 | 10 | 10 | 8 | 144 | 0.944444 | 4.34E-08 | 1 |
| **top 4K** | G2 brain cells | 10 | DUSP16 | MAPK10 | Q9BY84 | P53779 | 10 | 10 | 8 | 144 | 0.944444 | 4.34E-08 | 1 |
| **top 4K** | G2 brain cells | 11 | EWSR1 | MTMR9 | Q01844 | Q96QG7 | 10 | 10 | 8 | 144 | 0.944444 | 4.34E-08 | 1 |
| **top 4K** | G2 brain cells | 12 | GRB2 | PDCD6 | P62993 | O75340 | 10 | 10 | 8 | 144 | 0.944444 | 4.34E-08 | 1 |
| **top 4K** | G3 pituitary retina cells | 1 | UNC13A | YWHAB | Q9UPW8 | P31946 | 8 | 8 | 11 | 146 | 0.924658 | 4.56E-06 | 1 |
| **top 4K** | G3 pituitary retina cells | 2 | CHGB | SMAD2 | P05060 | Q15796 | 8 | 8 | 13 | 146 | 0.910959 | 9.48E-06 | 1 |
| **top 4K** | G3 pituitary retina cells | 3 | MAPK14 | SNAP91 | Q16539 | O60641 | 8 | 8 | 20 | 146 | 0.863014 | 7.37E-05 | 1 |
| **top 4K** | G3 pituitary retina cells | 4 | CTNNB1 | DUSP26 | P35222 | Q9BV47 | 7 | 8 | 6 | 146 | 0.833904 | 1.11E-05 | 1 |
| **top 4K** | G3 pituitary retina cells | 5 | JPH3 | RPS6KA3 | Q8WXH2 | P51812 | 7 | 8 | 6 | 146 | 0.833904 | 1.11E-05 | 1 |

| | | | | | | | | | | | | |
|---|---|---|---|---|---|---|---|---|---|---|---|---|
| **top 4K** | G3 pituitary retina cells | 6 | EPB41 | INA | P11171 | Q16352 | 7 | 8 | 9 | 146 | 0.813356 | 4.28E-05 | 1 |
| **top 4K** | G3 pituitary retina cells | 7 | ATCAY | UBA1 | Q86WG3 | P22314 | 7 | 8 | 10 | 146 | 0.806507 | 6.55E-05 | 1 |
| **top 4K** | G3 pituitary retina cells | 8 | PIK3R1 | SCG2 | P27986 | P13521 | 7 | 8 | 10 | 146 | 0.806507 | 6.55E-05 | 1 |
| **top 4K** | G3 pituitary retina cells | 9 | CALY | PRKCA | Q9NYX4 | P17252 | 7 | 8 | 11 | 146 | 0.799658 | 9.90E-05 | 1 |
| **top 4K** | G3 pituitary retina cells | 10 | HSP90AA1 | KCNH2 | P07900 | Q12809 | 7 | 8 | 12 | 146 | 0.792808 | 0.000141 | 1 |
| **top 4K** | G3 pituitary retina cells | 11 | ACTG1 | GRIA2 | P63261 | P42262 | 7 | 8 | 16 | 146 | 0.765411 | 0.000501 | 1 |
| **top 4K** | G3 pituitary retina cells | 12 | FYN | GABRG2 | P06241 | P18507 | 6 | 8 | 4 | 146 | 0.722603 | 5.61E-05 | 1 |
| **top 4K** | G4 immune vascular cells | 1 | EWSR1 | FLI1 | Q01844 | Q01543 | 12 | 15 | 16 | 139 | 0.684892 | 1.19E-05 | 1 |
| **top 4K** | G4 immune vascular cells | 2 | MAP3K3 | S100A8 | Q99759 | P05109 | 12 | 15 | 16 | 139 | 0.684892 | 1.19E-05 | 1 |
| **top 4K** | G4 immune vascular cells | 3 | CSNK2A2 | STATH | P19784 | P02808 | 11 | 15 | 8 | 139 | 0.675779 | 1.98E-06 | 1 |
| **top 4K** | G4 immune vascular cells | 4 | ITGA4 | TRIM28 | P13612 | Q13263 | 11 | 15 | 12 | 139 | 0.647002 | 1.78E-05 | 1 |
| **top 4K** | G4 immune vascular cells | 5 | LAPTM5 | SMURF2 | Q13571 | Q9HAU4 | 11 | 15 | 13 | 139 | 0.639808 | 2.45E-05 | 1 |
| **top 4K** | G4 immune vascular cells | 6 | CD48 | FYN | P09326 | P06241 | 10 | 15 | 5 | 139 | 0.630695 | 2.25E-06 | 1 |
| **top 4K** | G4 immune vascular cells | 7 | PIK3CG | RARA | P48736 | P10276 | 10 | 15 | 6 | 139 | 0.623501 | 4.69E-06 | 1 |
| **top 4K** | G4 immune vascular cells | 8 | CD53 | PRKCA | P19397 | P17252 | 10 | 15 | 9 | 139 | 0.601918 | 2.73E-05 | 1 |
| **top 4K** | G4 immune vascular cells | 9 | IKZF1 | RBBP8 | Q13422 | Q99708 | 10 | 15 | 9 | 139 | 0.601918 | 2.73E-05 | 1 |
| **top 4K** | G4 immune vascular cells | 10 | RHOH | TGFBR1 | Q15669 | P36897 | 10 | 15 | 9 | 139 | 0.601918 | 2.73E-05 | 1 |
| **top 4K** | G4 immune vascular cells | 11 | NCK1 | RAC2 | P16333 | P15153 | 10 | 15 | 11 | 139 | 0.58753 | 7.37E-05 | 1 |
| **top 4K** | G4 immune vascular cells | 12 | GRB2 | HCST | P62993 | Q9UBK5 | 10 | 15 | 12 | 139 | 0.580336 | 0.000119 | 1 |
| **top 4K** | G5 hematopoiesis | 1 | LMO2 | LYL1 | P25791 | P12980 | 7 | 7 | 12 | 147 | 0.918367 | 2.73E-05 | 1 |
| **top 4K** | G5 hematopoiesis | 2 | NEDD4 | NFE2 | P46934 | Q16621 | 6 | 7 | 2 | 147 | 0.843537 | 4.56E-06 | 1 |
| **top 4K** | G5 hematopoiesis | 3 | BTK | VAV1 | Q06187 | P15498 | 6 | 7 | 13 | 147 | 0.768707 | 0.000903 | 1 |
| **top 4K** | G5 hematopoiesis | 4 | LST1 | PRKCB | O00453 | P05771 | 5 | 7 | 11 | 147 | 0.639456 | 0.006053 | 1 |
| **top 4K** | G5 hematopoiesis | 5 | CD84 | PTPN11 | Q9UIB8 | Q06124 | 5 | 7 | 13 | 147 | 0.62585 | 0.009462 | 1 |
| **top 4K** | G5 hematopoiesis | 6 | PLEK | TGFBR1 | P08567 | P36897 | 5 | 7 | 14 | 147 | 0.619048 | 0.011658 | 1 |
| **top 4K** | G5 hematopoiesis | 7 | TAL1 | TCF3 | P17542 | P15923 | 4 | 7 | 1 | 147 | 0.564626 | 0.000708 | 1 |
| **top 4K** | G5 hematopoiesis | 8 | DPPA4 | VIM | Q7L190 | P08670 | 4 | 7 | 4 | 147 | 0.544218 | 0.005362 | 1 |
| **top 4K** | G5 hematopoiesis | 9 | GNAI2 | P2RY12 | P04899 | Q9H244 | 4 | 7 | 5 | 147 | 0.537415 | 0.008099 | 1 |
| **top 4K** | G5 hematopoiesis | 10 | MAPK14 | PLCB2 | Q16539 | Q00722 | 4 | 7 | 9 | 147 | 0.510204 | 0.026214 | 1 |
| **top 4K** | G5 hematopoiesis | 11 | SELPLG | SYK | Q14242 | P43405 | 4 | 7 | 12 | 147 | 0.489796 | 0.049667 | 1 |
| **top 4K** | G5 hematopoiesis | 12 | ELF1 | MLC1 | P32519 | Q15049 | 3 | 7 | 0 | 147 | 0.428571 | 0.003487 | 1 |

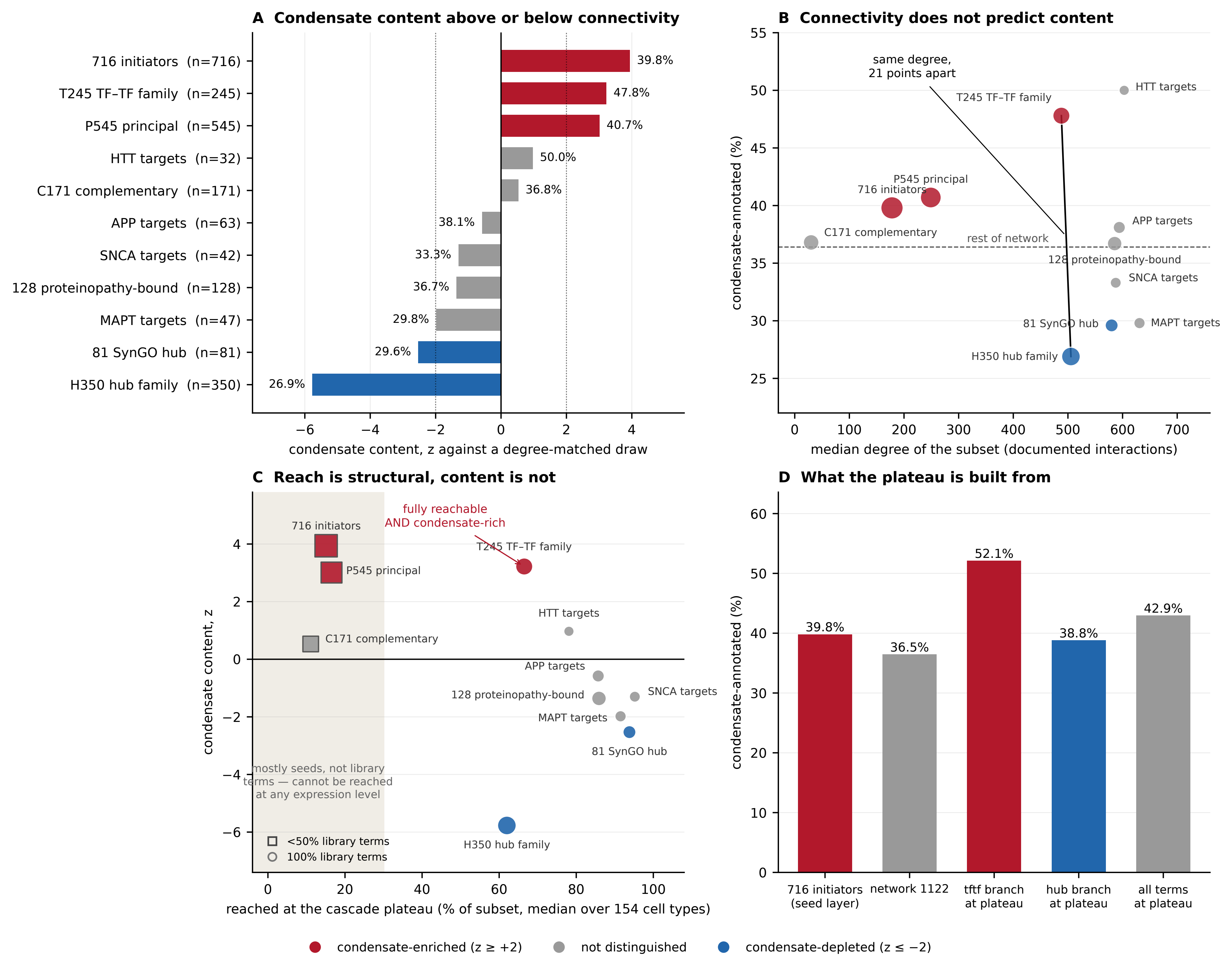

A Condensate content above or below connectivity
716 initiators (n=716)
T245 TF–TF family (n=245)
P545 principal (n=545)
HTT targets (n=32)
C171 complementary (n=171)
APP targets (n=63)
SNCA targets (n=42)
128 proteinopathy-bound (n=128)
MAPT targets (n=47)
81 SynGO hub (n=81)
H350 hub family (n=350)
39.8%
47.8%
40.7%
50.0%
36.8%
38.1%
33.3%
36.7%
29.8%
29.6%
26.9%
condensate content, z against a degree-matched draw
B Connectivity does not predict content
same degree, 21 points apart
HTT targets
T245 TF–TF family
P545 principal
716 initiators
APP targets
C171 complementary
rest of network
128 proteinopathy-bound
SNCA targets
81 SynGO hub
MAPT targets
H350 hub family
condensate-annotated (%)
median degree of the subset (documented interactions)
C Reach is structural, content is not
fully reachable AND condensate-rich
716 initiators
T245 TF–TF family
P545 principal
HTT targets
C171 complementary
APP targets
128 proteinopathy-bound
SNCA targets
MAPT targets
81 SynGO hub
mostly seeds, not library terms — cannot be reached at any expression level
<50% library terms
100% library terms
H350 hub family
condensate content, z
reached at the cascade plateau (% of subset, median over 154 cell types)
D What the plateau is built from
39.8%
36.5%
52.1%
38.8%
42.9%
716 initiators (seed layer)
network 1122
tftf branch at plateau
hub branch at plateau
all terms at plateau
condensate-annotated (%)
condensate-enriched (z ≥ +2)
not distinguished
condensate-depleted (z ≤ −2)

## Cell-type similarity in predicted PPI-pair repertoires

Numerical ≥1 availability · 154 cell types · 51,439 non-self pair types

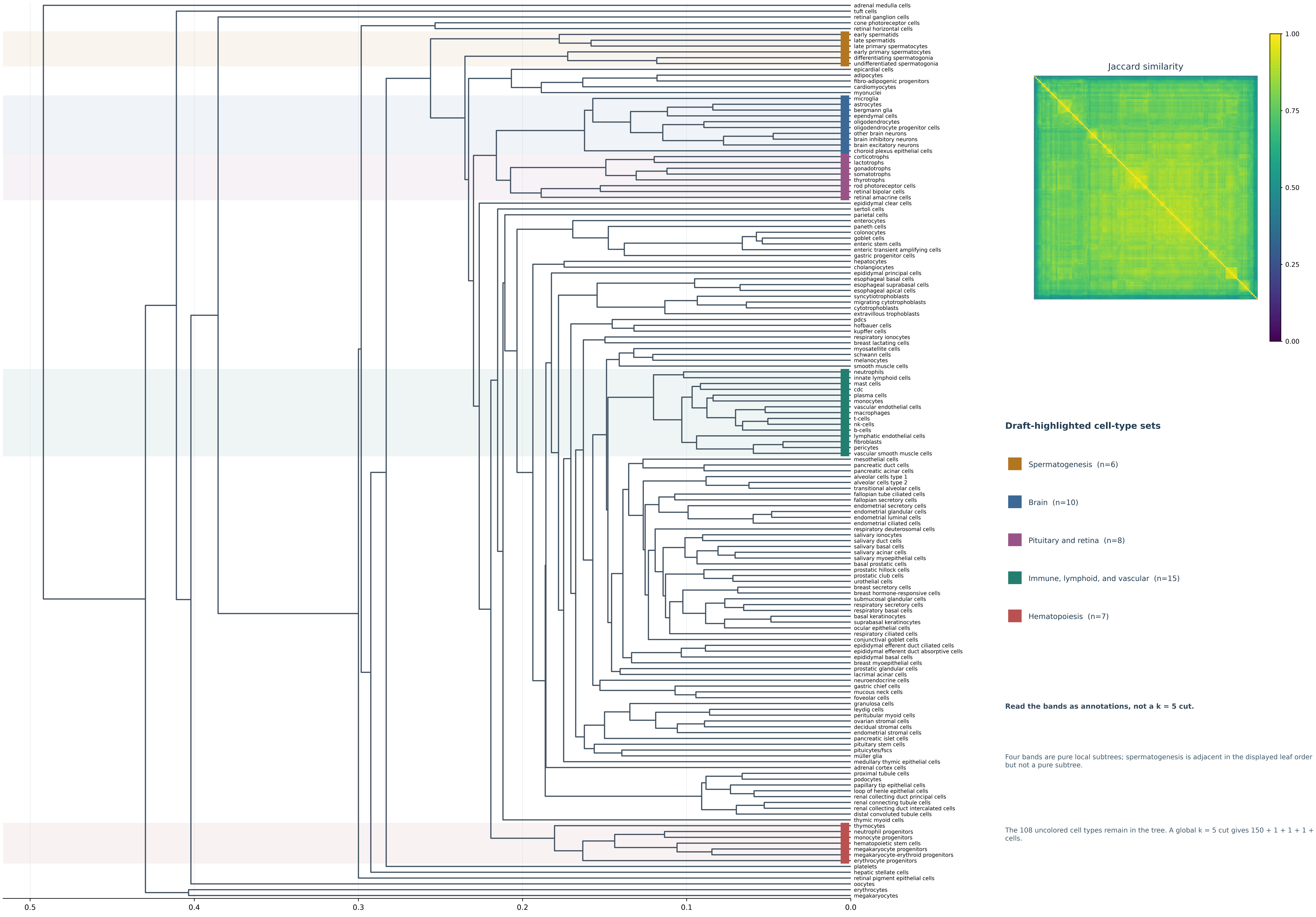

Figure. Unsupervised clustering of cell-type PPI-pair repertoires. Each of 154 cell types is represented by a binary profile of 51,439 non-self, unordered predicted pair types under the supplied numerical ≥1 availability rule. The dendrogram uses average linkage on Jaccard distance (1 minus intersection/union); optimal leaf ordering changes display order, not linkage. The adjacent heatmap shows cell-to-cell Jaccard similarity in the same leaf order (scale 0–1). Colored bands identify the five cell-type sets named in the draft (46 highlighted cells total); the other 108 cell types remain unassigned. Four sets are pure local subtrees in this tree. The six spermatogenesis cell types are adjacent leaves but do not constitute a single pure subtree. The five colors are interpretive annotations, not the five groups obtained by a global k = 5 cut, which yields one 150-cell group and four singleton outliers. Pair eligibility was inferred from expression availability and PPI-library term–member relationships, not experimentally verified dimer formation. HPA labels the underlying expression measurements nCPM; "numerical ≥1" follows the supplied cascade terminology.

# Condensate-associated interaction repertoires across human cell types

35 source-defined GO compartments · 154 HPA cell types · 51,439 reconstructed heterodimer types

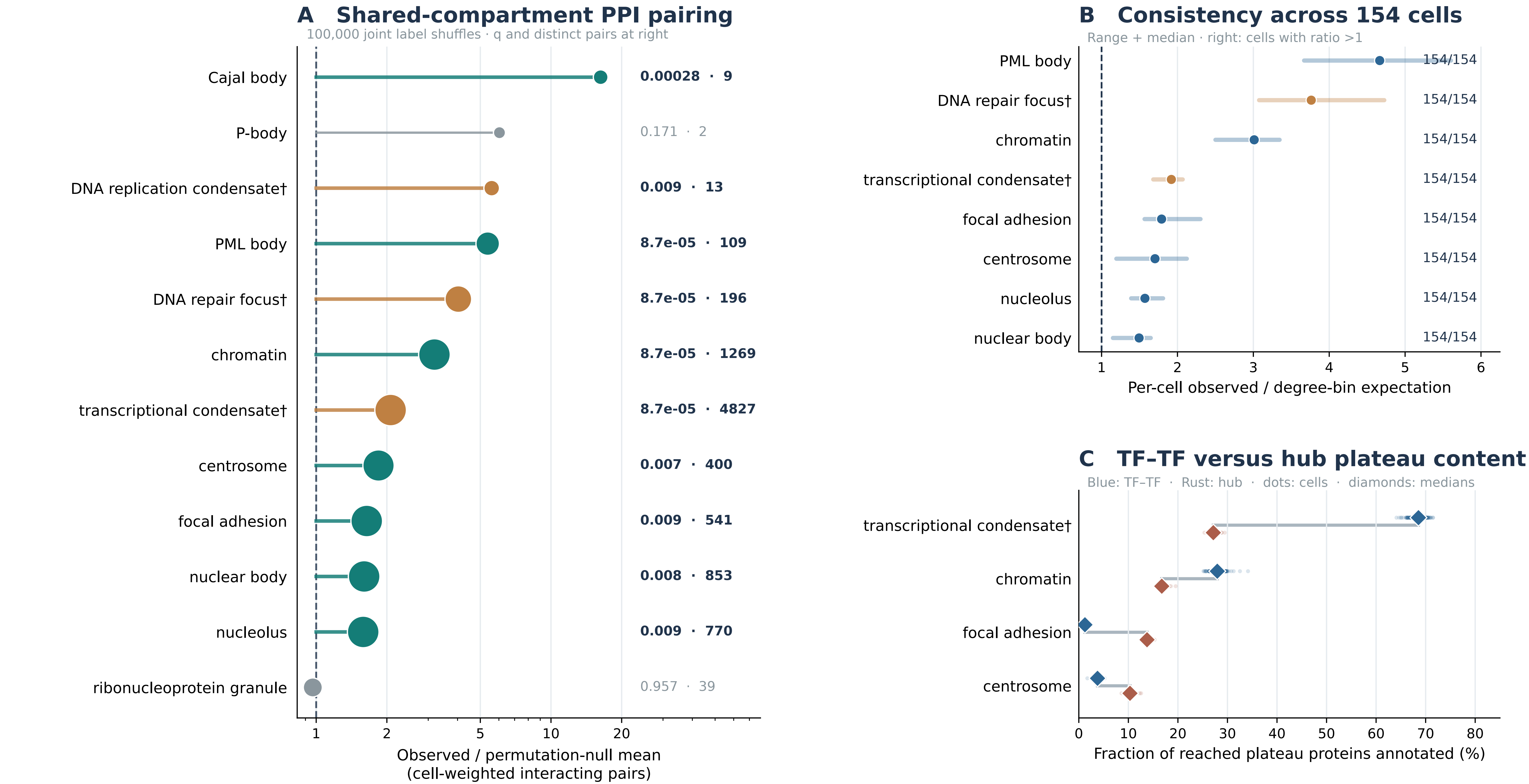

A q across 35 compartments; same 10 pass when each interaction is counted once. B per-cell ratios use a separate degree-only expectation (no per-cell p-values).

C protein content of reached branch plateaus, not pair assembly. † Functionally related GO proxy in source; annotations do not establish in-compartment dimer formation or phase separation.

Source: Banani et al., Developmental Cell (2022), Table S4I; HPA v25.1. Null: joint label shuffle within graph-degree × expression-breadth strata; BH across 35 terms.

# Pathogenic feature-overlap mutations in the PPI cascade

Banani et al. Table S4H × mapped cascade genes × 154-cell reached plateaus and interacting pairs

Independent self-assembly precedents: TP53 R337H ↓ tetramer stability · STAT3 Y640F ↑ homodimer stability

Neither allele-specific precedent establishes an effect on a cascade heterodimer or on condensate occupancy.

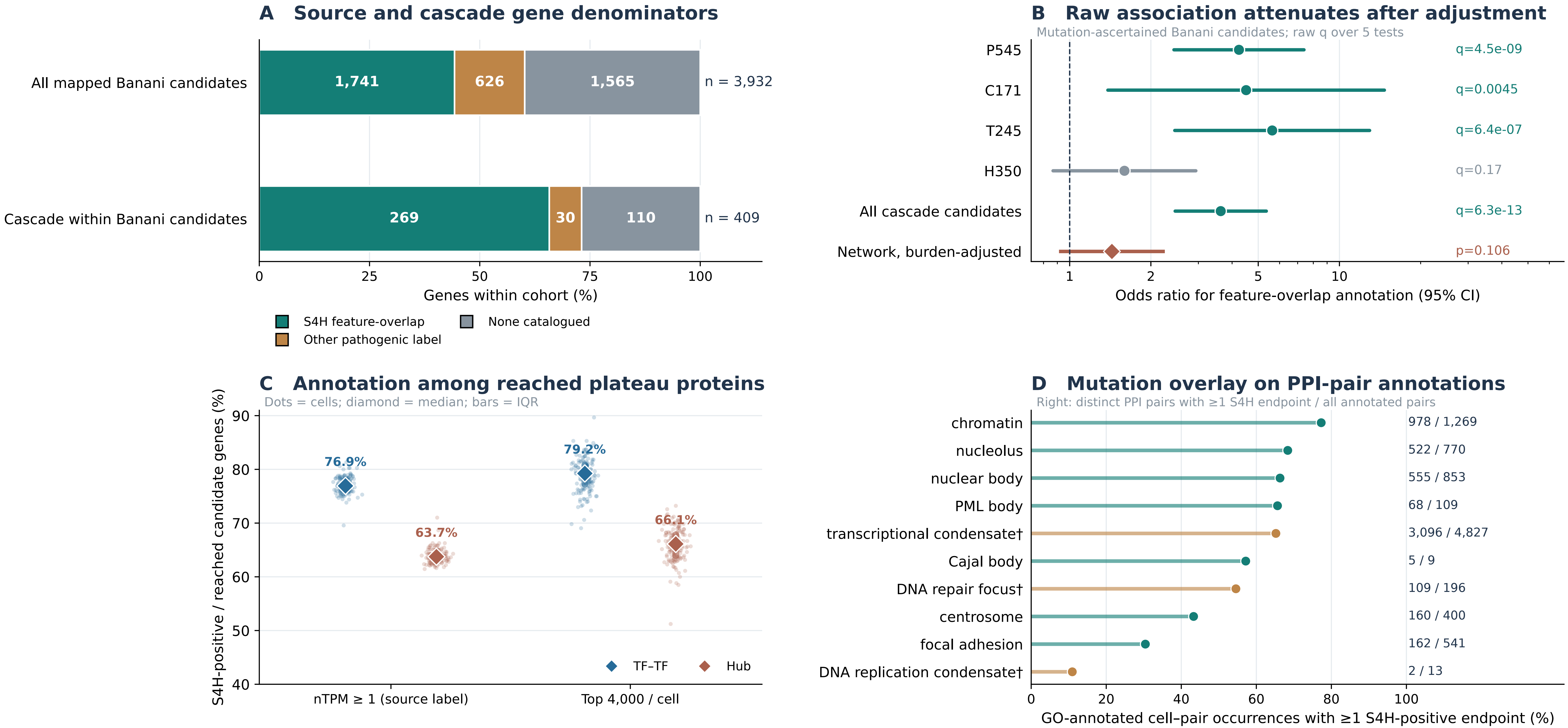

S4H marks pathogenic variants mapped to condensate-promoting features—not experimentally proven disruption. HPA cells are reference expression, not mutation carriers.

B The adjusted network comparison includes protein length, total pathogenic-mutation count and MID annotation. C–D are descriptive annotation overlaps, not variant-effect tests.

† Related functional GO proxy, not a matching compartment-localization term. Family sets overlap; source variant and PPI ascertainment remain limitations.

Source: Banani et al., Developmental Cell 57:1776–1788.e8 (2022), Table S4; HPA v25.1. PPI-pair GO enrichment tested separately over 35 terms.

# Curated LLPS evidence refines cascade pairing claims

PhaSePro / DrLLPS source roles × Banani feature-overlap mutations × 154-cell inferred PPI atlas

Driver and scaffold pairing exceeds a matched null; the client tier does not. Mutation associations attenuate after ascertainment adjustment.

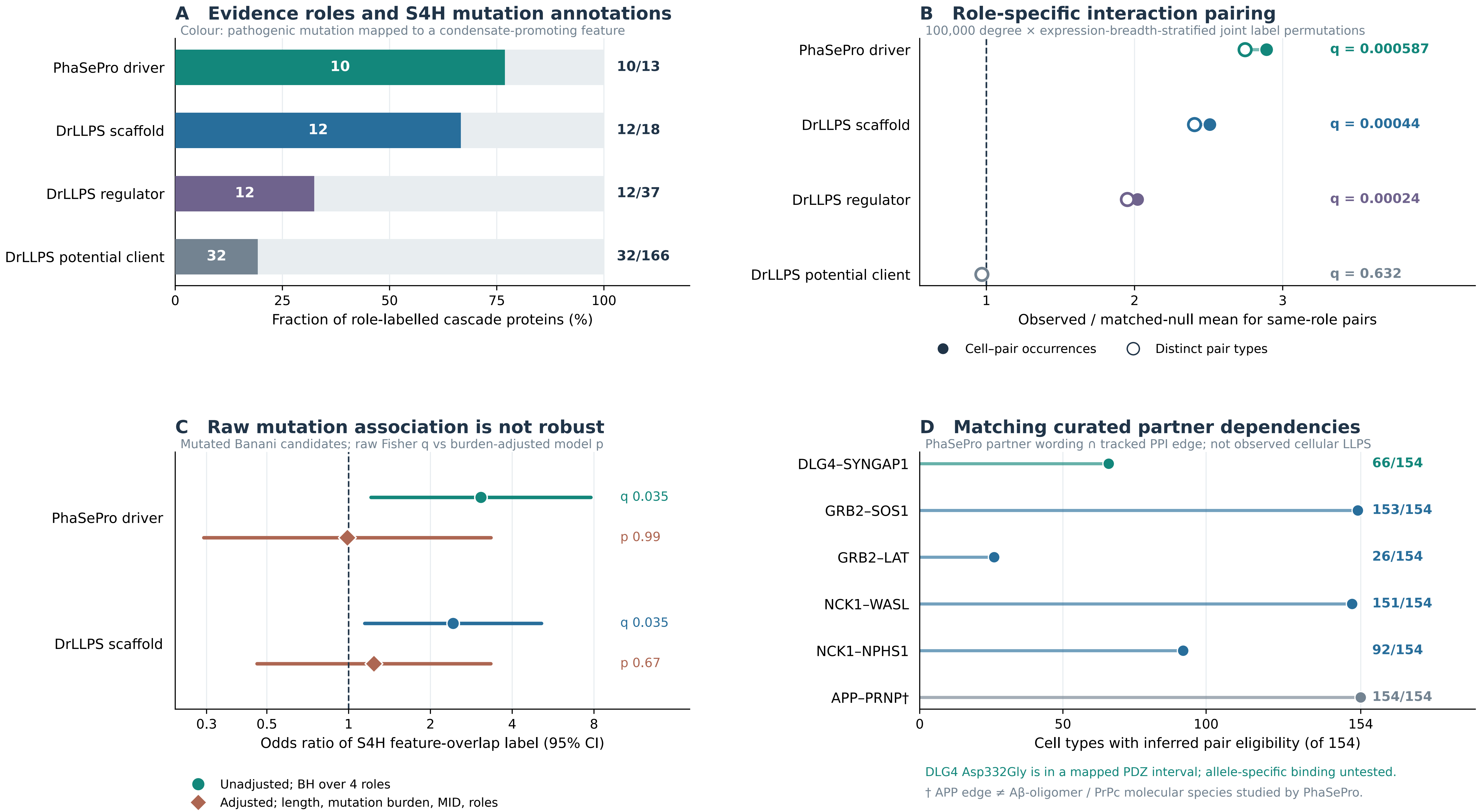

Roles are not interchangeable: DrLLPS potential clients are not validated LLPS drivers; regulators may modulate assembly without driving it.

Pair eligibility comes from reference HPA expression and the supplied PPI graph, not measured dimer formation or mutation-carrier cells.

Sources: PhaSePro (Mészáros et al., 2020); DrLLPS (Ning et al., 2020); Banani et al., 2022, Table S4; HPA v25.1.

# Synaptic condensate evidence converges on PSD-95

H350 cascade hubs × SynGO 2024 × DrLLPS/PhaSePro × tracked PPI pairs

Strongest mechanistic route: DLG4–SYNGAP1; DLG4–DLGAP1 is supported by multicomponent PSD reconstitution.

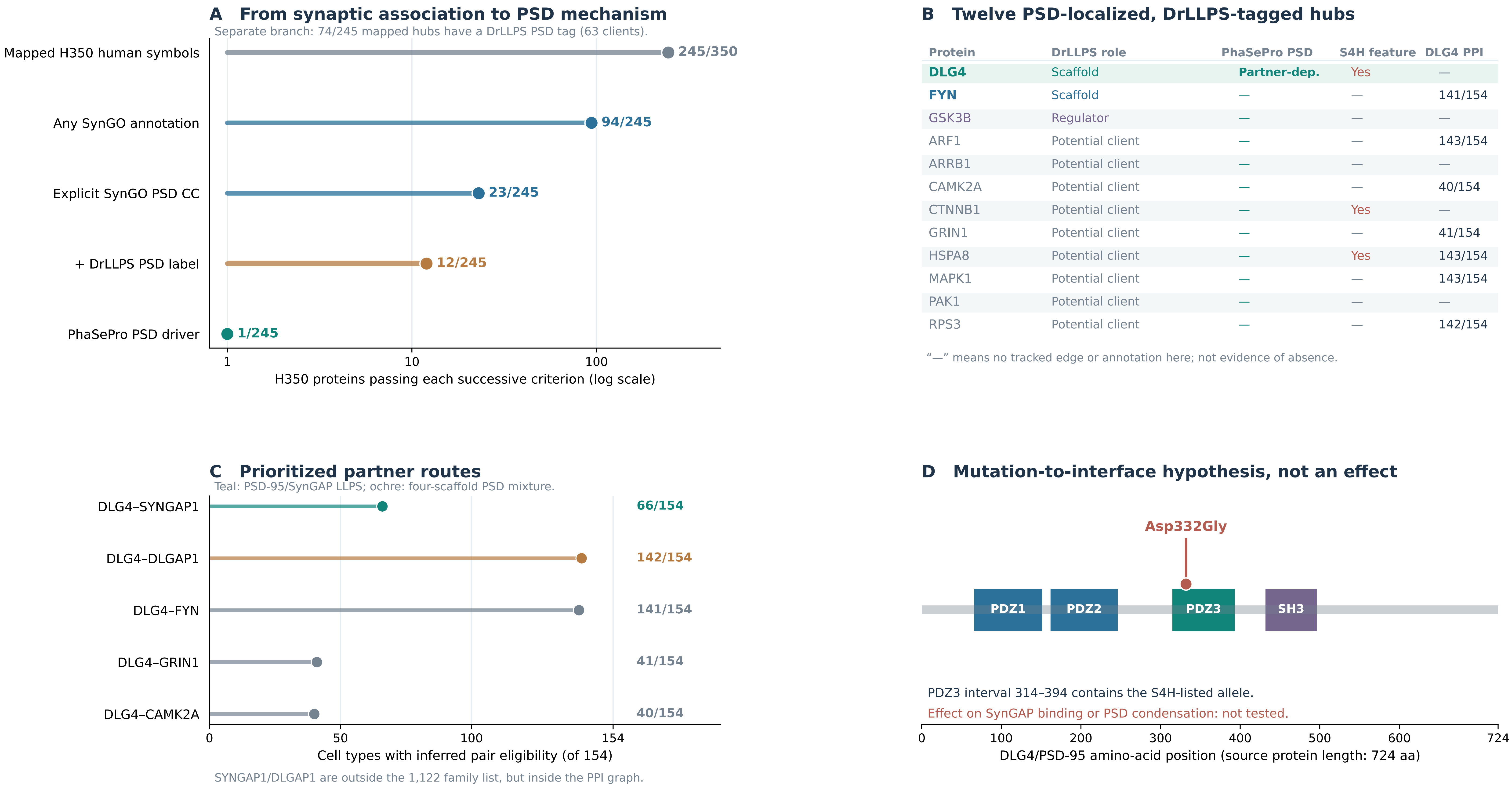

| Protein | DrLLPS role | PhaSePro PSD | S4H feature | DLG4 PPI |
|---|---|---|---|---|
| **DLG4** | Scaffold | **Partner-dep.** | Yes | — |
| **FYN** | Scaffold | — | — | 141/154 |
| GSK3B | Regulator | — | — | — |
| ARF1 | Potential client | — | — | 143/154 |
| ARRB1 | Potential client | — | — | — |
| CAMK2A | Potential client | — | — | 40/154 |
| CTNNB1 | Potential client | — | Yes | — |
| GRIN1 | Potential client | — | — | 41/154 |
| HSPA8 | Potential client | — | Yes | 143/154 |
| MAPK1 | Potential client | — | — | 143/154 |
| PAK1 | Potential client | — | — | — |
| RPS3 | Potential client | — | — | 142/154 |

SynGO location and DrLLPS association are distinct from a demonstrated LLPS driver; all pair counts are inferred from reference expression.

Neither DLG4 variant carriage nor condensate occupancy was measured across the 154 cell types. Missing PPI edges do not refute reconstituted PSD assemblies.

Sources: SynGO 2024 (Harmonizome); PhaSePro; DrLLPS; Banani et al. (2022), S4H; Zeng et al., Cell (2016, 2018).

# Candidate condensate-feature variants relative to partner-bound interfaces

Experimentally resolved fragments; reference mutation sites transferred to UniProt numbering

Blue: target Cα trace · Green: partner · Cyan/gold: ≤5 Å interface residues · Colored large dots: S4H mutation sites

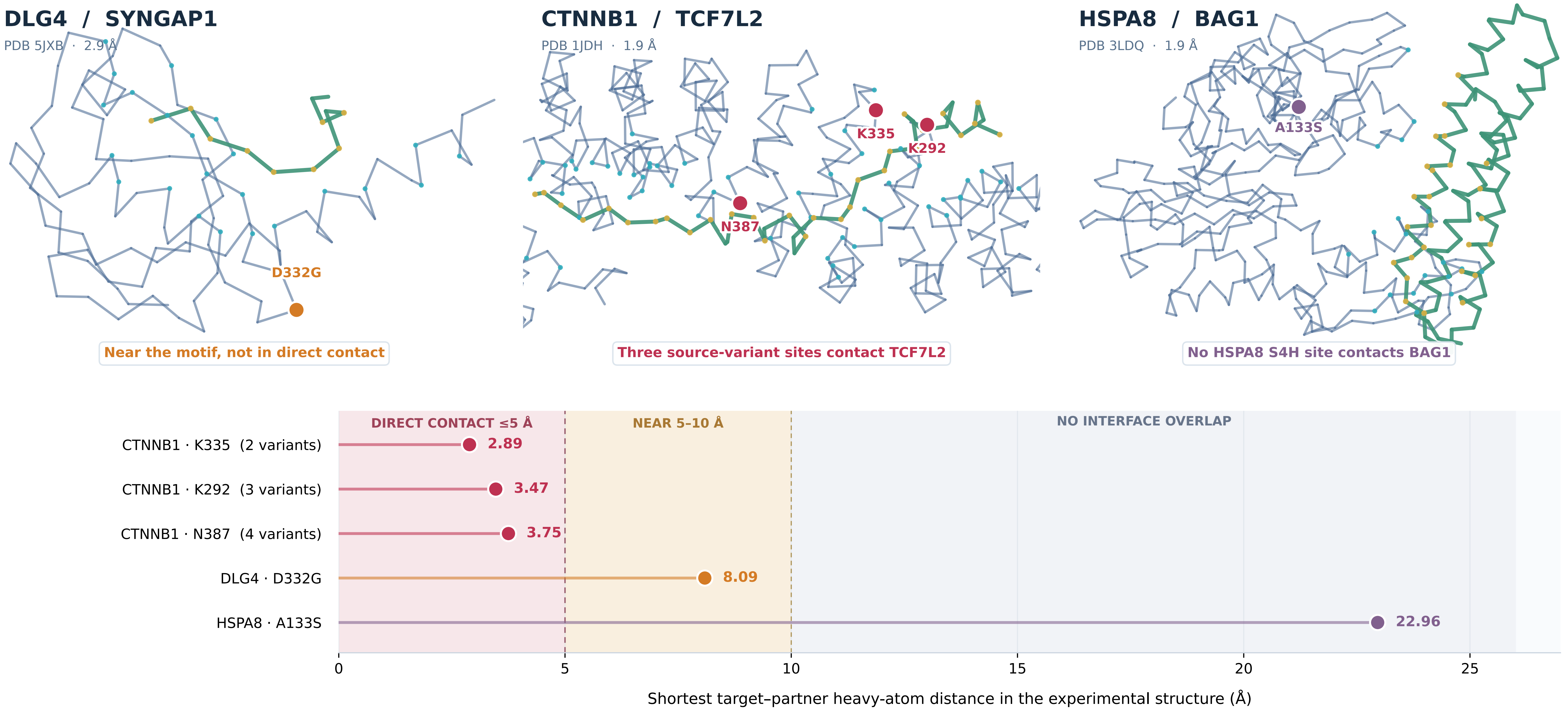

Cα traces are orthographic projections of PDB coordinates; distances are computed in 3D from heavy atoms, not from the image. Only one partner interface per target is examined.

5JXB is a human DLG4–mouse SYNGAP motif fusion displayed as two chains; 1JDH and 3LDQ are bound human fragments. Feature overlap and geometric proximity do not establish disrupted binding.